\documentclass[12pt,preprint]{aastex}

\begin{document}
 
\newcommand{\kms}{km s$^{-1}\;$}
\newcommand{\msun}{M_{\odot}}
\newcommand{\rsun}{R_{\odot}}
\newcommand{\rat}{$R_{2}$~}
\newcommand{\fbss}{$F_{BSS}^{HB}$}
\newcommand{\mav}{\langle M_{HB} \rangle}
 
\title{Variable Stars in the Open Cluster NGC 7142}

\author{Eric L. Sandquist, Andrew W. Serio\altaffilmark{1}} 
\affil{San Diego State University, Department of Astronomy, San Diego,
CA 92182}
\email{erics@mintaka.sdsu.edu,aserio@gemini.edu}
\altaffiltext{1}{Current address: Gemini Observatory, Southern Operations 
Center, AURA, Casila 603, La Serena, Chile}

\author{Matthew Shetrone}
\affil{University of Texas, McDonald Observatory, HC75 Box 1337-L 
    Fort Davis, TX, 79734}
\email{shetrone@astro.as.utexas.edu}

\begin{abstract}
We present new discoveries of variable stars near the turnoff of the old
open cluster NGC 7142. Contrary to previous studies, we detect
eight contact or near contact eclipsing binaries (including three near the
cluster turnoff), and most of these have good probability of being cluster
members. We also identified one long-period variable that resides far to the
red of the cluster giant branch, and four new detached eclipsing binaries.

We have re-examined the question of distance and reddening for the cluster,
and find that the distance is larger and reddening lower than in most previous
studies. In turn this implies that NGC 7142 is probably slightly younger than
M67, about 3 Gyr old. With an age of this size, NGC 7142 would be one of a
small group of clusters with main sequence turnoff stars at the transition
between convective and radiative cores.
\end{abstract}


\section{Introduction}

Star clusters have long been testbeds for our understanding of the evolution
of stars, the physics of the gas in their interiors, and the dynamics of the
member stars. Variable stars can provide a great deal of additional
information that allows us to tighten the screws on theoretical models
attempting to predict cluster stellar characteristics from first
principles. Pulsating stars can provide information about internal structure
and stellar mass. Short-period semi-detached and
contact binaries may contain encoded information about the dynamical history
of the cluster. But perhaps most valuable of all are detached eclipsing
binaries because they can provide high precision measurements of masses and
radii for individual stars with a minimum of theoretical interpretation.

Variability studies of NGC 7142 have previously been carried out by \citet{ct}
and \citet{rh}.  \citeauthor{rh} searched for low-amplitude variables on a
single night of observations, finding eight stars with significant light
variations. \citeauthor{ct} expected to find a number of short-period variable
stars based on discoveries in clusters like NGC 188.  V375 Cep was the only
variable star they discovered in the cluster, although they lacked sufficient
observations to determine a period. The variable is of interest because its
photometry puts it very close to the turnoff of the cluster's color-magnitude
diagram, within the main sequence band. Masses for these stars would be
valuable constraints on the cluster age, but in addition, one of the two stars
could be significantly evolved from the zero-age main sequence. This means
that the radius of the star could help refine the age of the cluster.  [One
  other semi-regular variable, V582 Cep, was known \citep{gsc} to be in the
  field of the cluster, but was outside the field observed by us.

In one of the earliest studies of the cluster, \citet{vh} found that NGC 7142
was intermediate in age between NGC 7789 and M67, approximately $1.5-4$
Gyr. \citet{ct} found that it was intermediate in age between M67 and NGC
188, or approximately $4-5$ Gyr.  \citet{carr} estimated an age of 4.9 Gyr by
comparing synthetic color-magnitude diagrams (generated from model isochrones)
with photometry. \citet{saloc} give an age of $4\pm1$ Gyr for NGC 7142 based
on the magnitude difference between the subgiant branch and the red giant
clump. Recently though, \citet{janes} derived a much larger age of
$6.9\pm0.8$ Gyr, using synthetic CMDs to match the turnoff color and red giant
clump color and magnitude. Much of the disagreement may be the result of
differential extinction in the cluster and the small numbers of evolved stars,
making the reference points in the CMD difficult to identify.

In order to improve the precision of the age determination for this old open
cluster (and others), we have undertaken a program to identify and analyze
age-sensitive eclipsing binary stars. This paper presents the results of our
variable search, and a following study presents and analysis of the most
promising candidates.

\section{Observational Material and Data Reduction}\label{obs}

\subsection{Photometry}

All of the photometry for this study was taken with the Mount Laguna
Observatory 1 m telescope using a camera having a 2048$\times$2048 pixel CCD
and approximately a $13\farcm5\times13\farcm5$ field of view. Table
\ref{phottab} lists the nights during which images were taken for this
article. 17 nights of data were taken in $R_C$ band for the purpose of
identifying variable stars in the cluster, and subsequent nights were taken
primarily to refine the ephemerides of the most important variables and to
determine light curves in different filter bands.

Differential photometry was undertaken using the image subtraction package
ISIS \citep{isis}. Our procedure is very similar to that of \citet{talaman},
so for most details, we refer the reader to that paper.  Our
image sets in $B$, $V$, $R_C$, and $I_C$ filters were interpolated to a common
image coordinate system, and then processed separately by filter. 

The output of ISIS is a difference flux measured on the subtracted images.  To
convert these difference fluxes into magnitudes, star fluxes were measured on
the reference frame. ISIS's algorithm is a modified aperture photometry
routine that employs the reference image point-spread function (PSF) for
weighting purposes. The reference PSF is first transformed to the seeing of
the image under consideration. Only the portion within an aperture of {\tt
  radphot} pixels will be used, but it is normalized to a larger aperture of
{\tt rad\_aper} pixels. The pixel values in the subtracted image are weighted
by this transformed PSF. In this study, we used {\tt radphot = 4} pix and {\tt
rad\_aper = 10} pix.

%This procedure brings up another potential issue: because fixed apertures are
%being used in the photometry, a seeing-dependent error may be introduced.
%Other researchers \citep{bald,gillon} have found that ISIS typically produces
%light curves with amplitudes that are systematically too small. Because open
%clusters like NGC 7142 typically have some variable stars in relatively
%uncrowded parts of the field, it is possible to compare the ISIS results to
%more conventional aperture photometry with a curve of growth analysis
%\citep{}. By this means, we verified that the underestimation of variability
%amplitudes by ISIS occurred here as well. In addition, we identified
%systematic trends in the ISIS light curves that did not appear in the aperture
%photometry.

We discovered a bug in the determination of uncertainties in the difference
fluxes derived by ISIS. As part of the image subtraction process, the overall
flux scaling from frame to frame is fit and corrected in ISIS. This scaling
affects the size of the noise and therefore the flux uncertainties, but this
was not done in the last version of ISIS. This had the unfortunate effect of
depressing computed uncertainties on measurements with low fluxes. We fixed
this error in our copy of the ISIS code (version 2.2), and all of the results
here use this corrected version.

\subsubsection{Photometric Calibration}

We took calibration images under photometric conditions on the night of 25
October 2008. We observed the standard fields PG0231+051, SA 92, SA95, SA 98,
and NGC 6940, and used standard values taken from \citet[][retrieved August
2009]{stet}. The standard fields were observed between 3 and 10 times per
filter in $BVI$ at airmasses that ranged from 1.034 to 1.998. All together
this resulted in more than 4000 standard star observations per filter covering
a color range $-0.5 \la (B-I) \la 5$.

\begin{figure}
% clusters/cal/mlooc/rsdnet.
\plotone{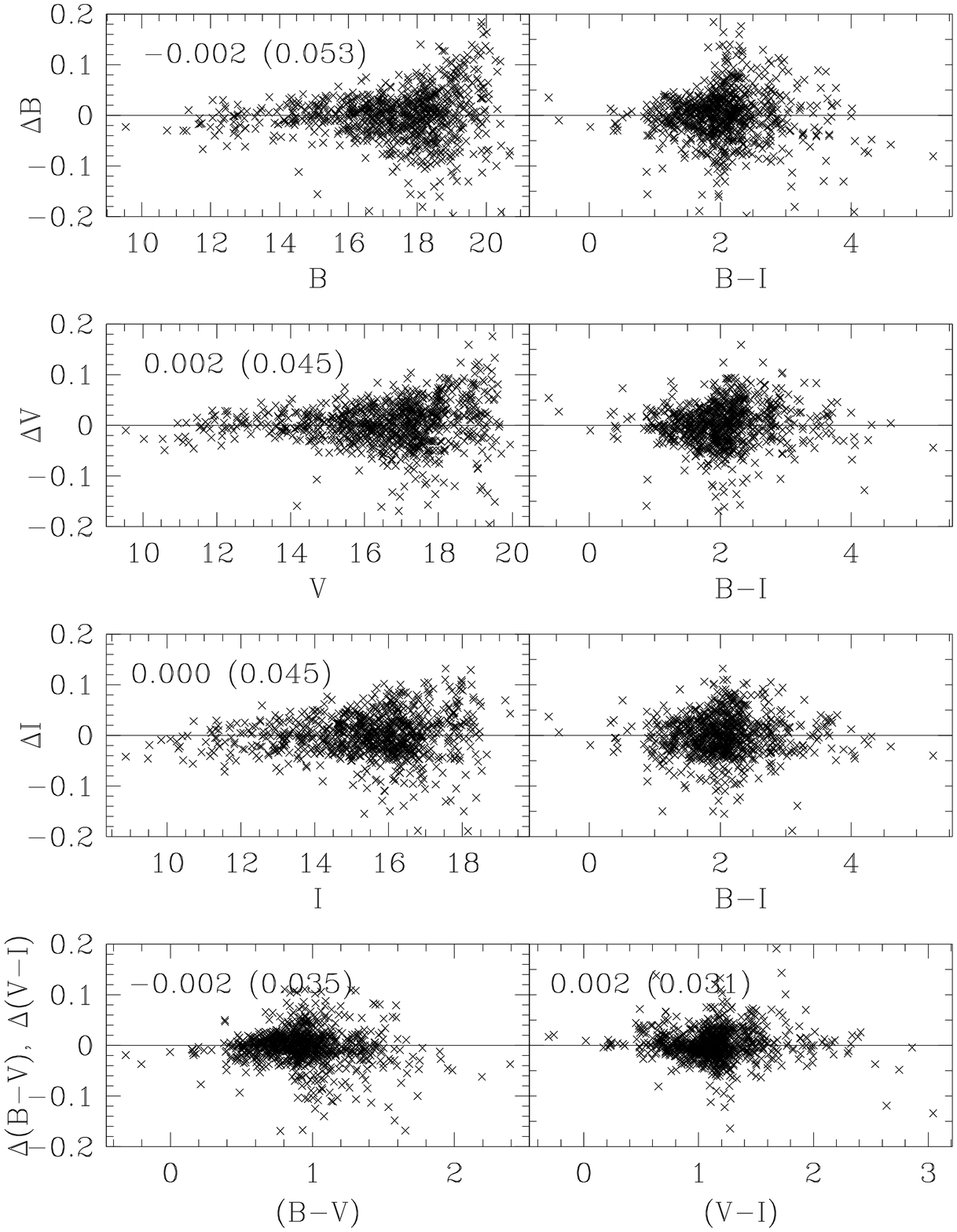}
\caption{Residuals between our calibrated photometry of photometric standards
  and the standard magnitudes of \citealt{stet} (in the sense of this study
  minus Stetson's).\label{stetcomp}}
\end{figure}

We derived aperture photometry from all frames using DAOPHOT, and made curve
of growth corrections using the program DAOGROW.
The observations were transformed to the standard system using the
following equations:
\[ b = B + a_0 + a_1 (B-I) + a_2 X \]
\[ v = V + b_0 + b_1 (B-I) + b_2 X \]
\[ i = I + c_0 + c_1 (B-I) + c_2 X \]
where $b, v,$ and $i$ are instrumental magnitudes, $B, V$, and $I$ are
standard-system magnitudes, $X$ is airmass, and $a_i, b_i$, and $c_i$
are coefficients determined from least-squares fits. Fig. \ref{stetcomp}
shows the residuals of the comparison of our standardized observations
and the Stetson values.

Clusters to be calibrated were observed with a range of exposures times on the
same night. For NGC 7142, there were 9 observations in $B$ (1$\times$60 s,
5$\times$120 s, and 3$\times$300 s), 8 observations in $V$ (3$\times$60 s,
2$\times$120 s, and 3$\times$300 s), and 11 observations in $I$ (4$\times$10
s, 2$\times$60 s, 2$\times$120 s, and 3$\times$300 s).

\begin{figure}
% clusters/cal/mlooc/ctcomp.
\plotone{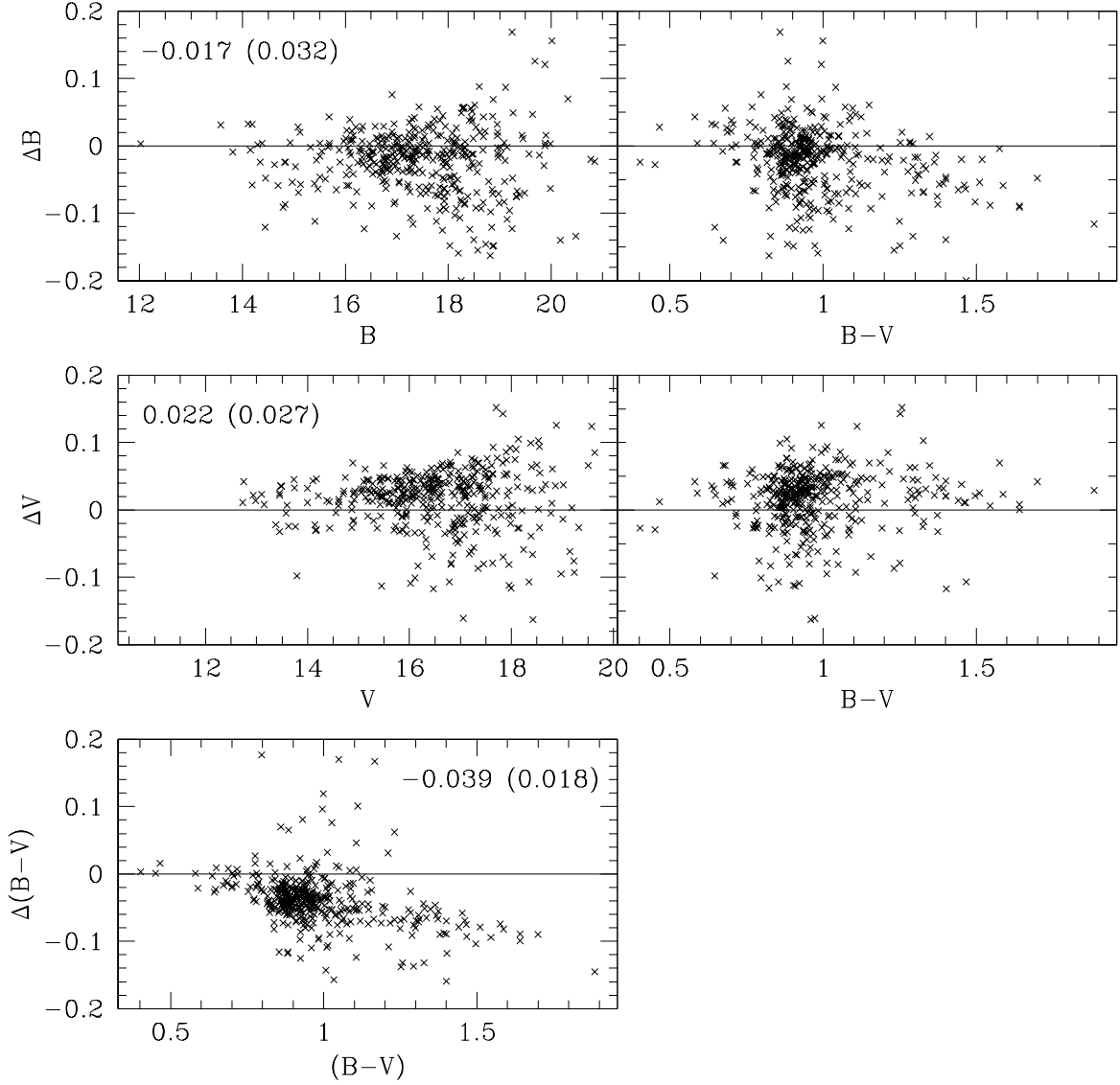}
\caption{Residuals between our photometry of NGC 7142 and that of
  \citet{ct}, in the sense of this study minus theirs. The quoted numbers are
  the median residual and the semi-interquartile range in $B$, $V$, and $B-V$,
  respectively.\label{ctcomp}}
\end{figure}

\begin{figure}
% clusters/n7142/isos/janescomp.
\epsscale{0.8}
\plotone{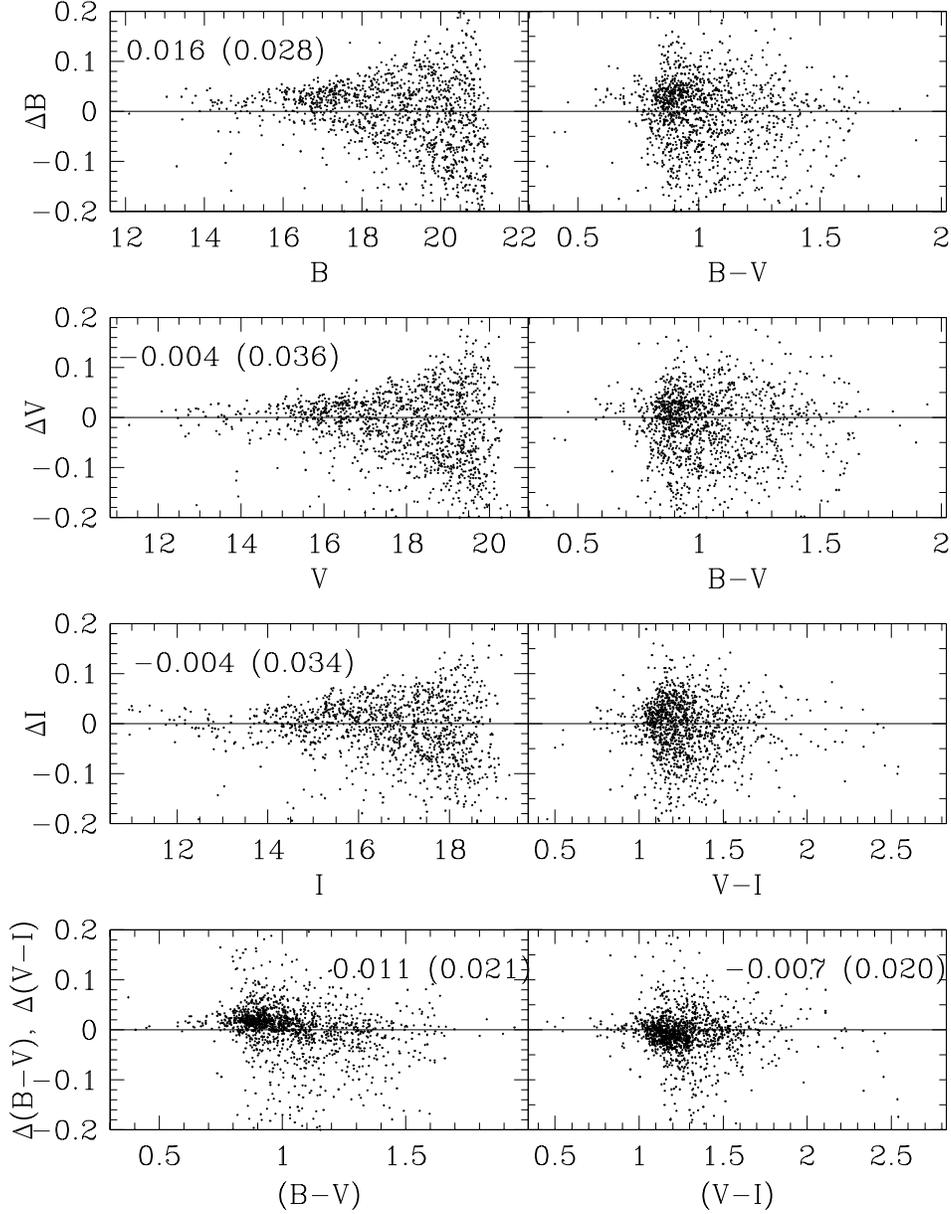}
\caption{Residuals between our photometry of NGC 7142 and that of
  \citet{janes}, in the sense of this study minus theirs. The quoted numbers
  are the median residual and the semi-interquartile range. The median
  magnitude residuals were calculated based on stars with $B < 18$, $V < 18$,
  and $I < 17$, respectively.\label{janescomp}}
\end{figure}

Little attention had been paid to NGC 7142 in the past, with \citet{ct}
providing the most extensive set of photometry. Fig. \ref{ctcomp} shows a
comparison of our photometry with theirs. There are small zeropoint
differences between the two studies, but more importantly, there is a clear
trend with color. We believe our calibration is superior based on number of
standards and improved technique. We have also compared to the more recent
photometry by \citet{janes}. The zeropoint differences are again small as
shown in Fig. \ref{janescomp}, but there are also no obvious trends with
color.

\begin{figure}
% clusters/cal/mlooc/cmdcal.n7142.
\plotone{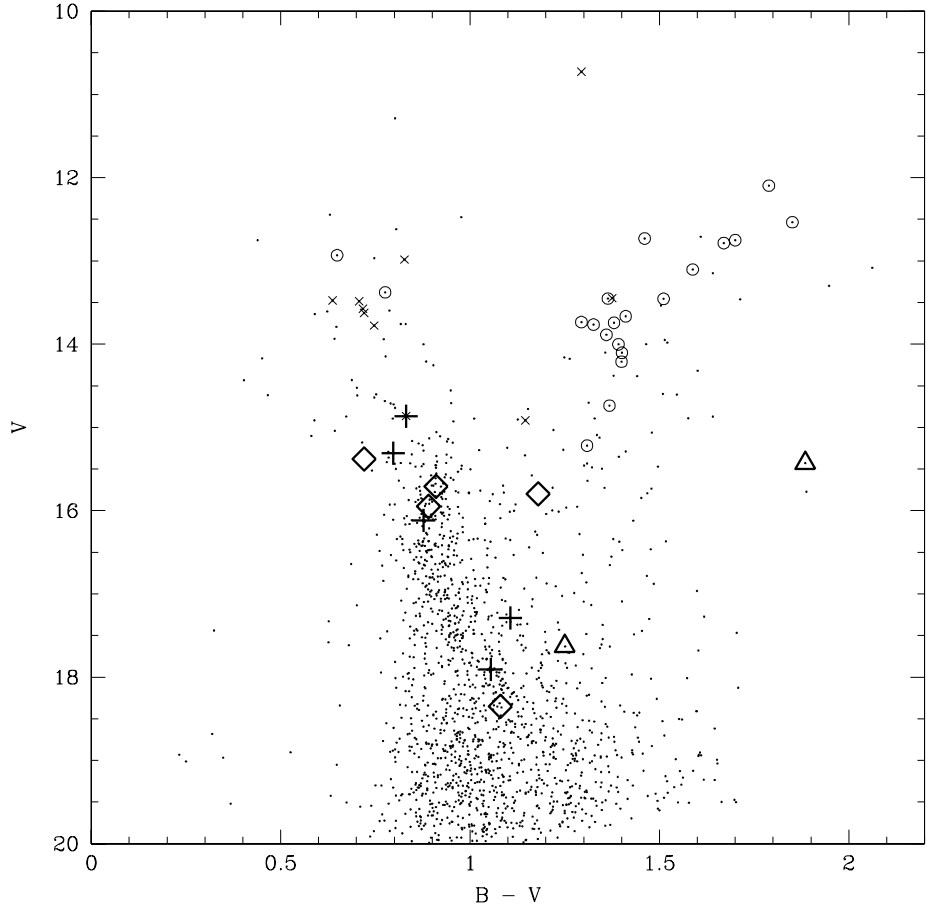}
\caption{Color-magnitude diagram for NGC 7142 with variables identified.
  Detached eclipsing binaries are shown with crosses, contact and near contact
  binaries are shown with diamonds, and quasi-periodic and irregular variables
  are shown with triangles. Probable cluster members (identified from
  spectroscopic radial velocities) are shown with small open circles, and
  nonmembers are shown with small $\times$.\label{cmd}}
\end{figure}

The color-magnitude diagram for our field is shown in Fig. \ref{cmd}. Note
that much of the scatter probably comes from differential reddening resulting
from the cluster's proximity on the sky to the reflection nebula NGC 7129
(although the cluster is well behind the cloud). However, to help identify the
cluster's sparsely populated giant branch, we identified all the stars with
radial velocities from spectra that have been presented in the literature
\citep{fj,jaco1,spec}.

\subsection{Spectra}
% currently identical to NGC6819 paper - refer to that?

Spectra for three variable star targets and three red clump star candidates
were obtained at the Hobby-Eberly Telescope (HET) with the High Resolution
Spectrograph (HRS, \citealt{tull}) as part of normal queue scheduled observing
\citep{shetet}. The observed stars are listed in Table \ref{spectab}. HRS was
configured to HRS\_30k\_central\_316g5936\_2as\_0sky\_IS0\_GC0\_2x3 or
HRS\_30k\_central\_600g5822\_2as\_2sky\_IS0\_GC0\_2x3 to achieve R=30,000
spectra covering 4100\AA\ to 7800\AA\ or 4825\AA\ to 6750\AA\, respectively.
Exposure times ranged from 600 seconds to 1200 seconds.  The signal-to-noise
was typically 25 per resolution element at 5800 \AA.

The spectra were reduced with IRAF ECHELLE scripts.  The standard IRAF scripts
for overscan removal, bias subtraction, flat fielding and scattered light
removal were employed.  For the HRS flat field we masked out the Li I, H I and
Na D regions because the HET HRS flat field lamp suffered from faint emission
lines.  The spectra were combined into a single long spectrum for the blue and
red chips.  Radial velocities were determined from cross-correlation using the
IRAF task {\tt fxcor} using the solar spectra.  The heliocentric correction
was made using the IRAF task {\tt rvcorrect}. The radial velocities are
discussed in more detail in \S \ref{debs}.

\section{Variable Stars}

We present the detected variable stars in Table \ref{vartab}. We have started
a new nomenclature for all detections (except the one known variable V375
Cep), ordered by $V$ magnitude. \citet{rh} identified 8 stars as ``suspected''
variable stars and an additional 12 as ``potential'' (less likely)
variables. Even though all of their systems were in our field and all but five
were faint enough to be within the dynamic range of our images, only two of
the stars in their list were identified as variables in our observations.

\subsection{Contact and Near-Contact Binaries}

Because our calibration observations were taken over two relatively short
periods on one night, we determined corrections to the calibrated magnitudes
for each of our close binary systems to give us values at maximum light. These
corrections were as large as about 0.1 mag in a few cases, which is enough to
affect judgements about cluster membership from CMD position.

Short-period contact and near-contact binary star systems have only been found
in open clusters older than about 600 Myr \citep{rucin,zhang}. For fainter
systems, this comes about as a result of the long timescales for tidal
interactions to produce orbital decay. At the bright end though, the evolution
of one component of a binary can accelerate the process. NGC 7142 has 4
contact or near-contact systems near the cluster turnoff, comparable in number
to systems fainter on the main sequence. Similar groupings of systems are seen
in old open clusters like M67 \citep{ssm67} and NGC 7789 \citep{jahn},
supporting the idea that mass transfer (and potentially coalescence of the
stars) can produce blue straggler stars. This does not appear to be universal
among open clusters, however. \citet{rucin}, for example, found that such
systems are not concentrated at the turnoff, but can be found spread
brightward (among blue stragglers) or faintward (along the main sequence).
% Mateo Nemec NGC 5466?

Before drawing conclusions, we can assess whether the contact binaries are
likely to be members using the period-luminosity-color (PLC) relationship from
\citet{ruc97}
\[ M_V = -4.44 \log P + 3.02 (B-V)_0 + 0.12 \] %0.22 mag sigma
and comparing the implied distance modulus to what is estimated from
isochrones. The significant differential reddening in NGC 7142 will reduce the
sharpness of this tool by introducing additional random scatter of
approximately 0.15 mag, but in some cases we should be able to make strong
statements. Based on our later discussion of the red giant clump (\S
\ref{dist} and \ref{aged}), we find that the mean reddening for the cluster
appears to be $E(B-V) \approx 0.36$, with a distance modulus $(m-M)_V =
13.18$. V5 has an implied distance modulus about 2 mag smaller [$(m-M)_V =
  11.1$], and V13 has a value almost a magnitude larger (14.0).  Two of the
remaining systems are very consistent with cluster membership (V3: 13.1; V9:
13.0), while two others are possible members (V7: 13.5; V12: 13.4). Because
NGC 7142 is similar in age to other clusters (such as M67) known to have
contact binary members, these new detections are understandable, although the
systems don't seem to show any sign of being concentrated toward the cluster
center.

% V6 [$(m-M)_V = 12.35$]. 

%V3 (m-M)_V = 13.13
%V5           11.11
%V6 (EB)      12.56
%V7           13.52
%V9           13.02
%V12          13.43
%V13          13.98
%V14 not calibrated

\begin{figure}
% clusters/n7142/ew
\plotone{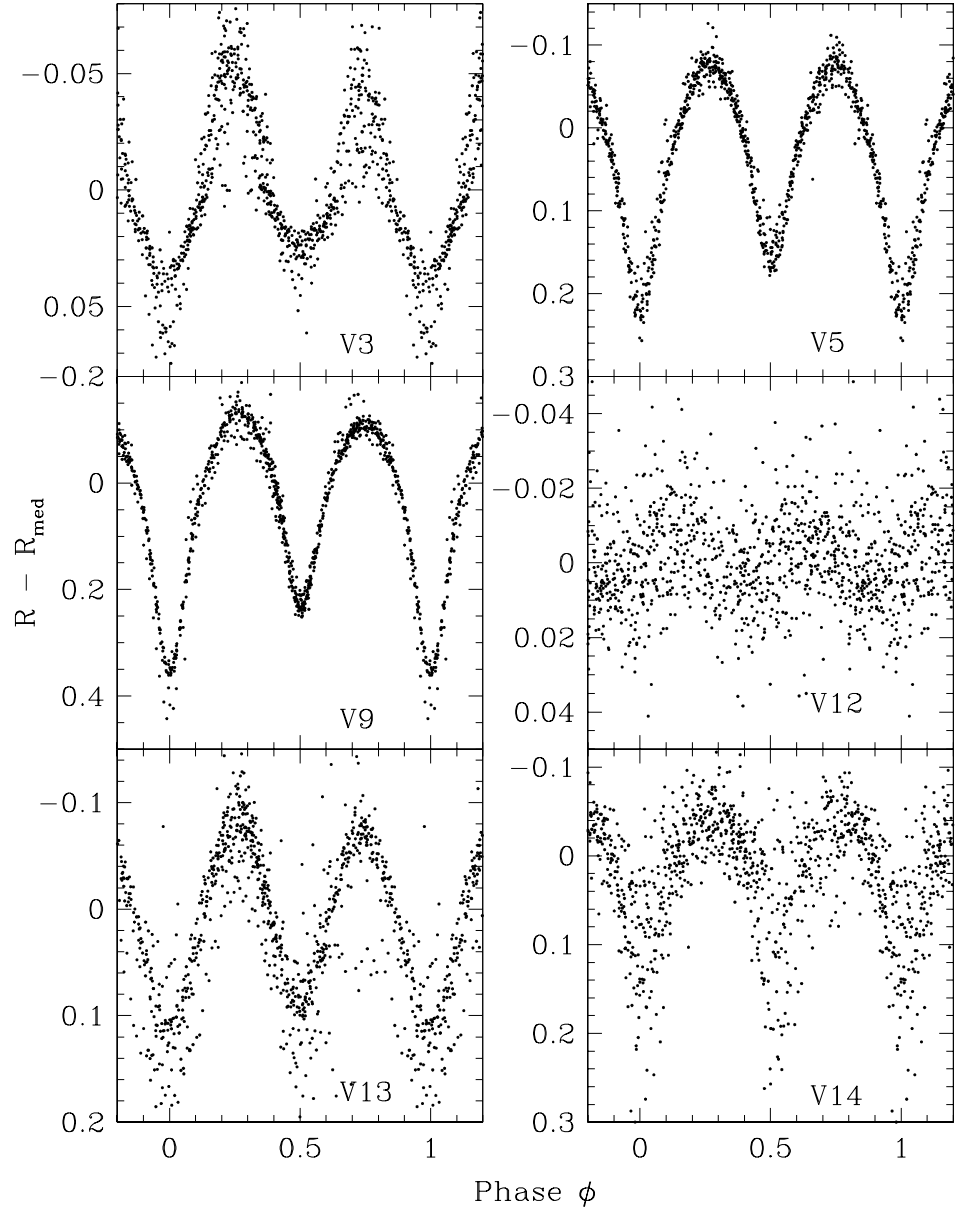}
\caption{$R_c$ phased light curves for short-period binaries V3, V5, V9, V12,
  V13, and V14.
\label{lcwuma}}
\end{figure}

{\bf V3 (Fig. \ref{lcwuma}):} This is one of the systems that was
``suspected'' to be variable by \citet{rh} that we confirm as variable (their
star 279). Based on the PLC relation, this system may be a cluster blue
straggler. There are slight but distinct differences between the depths of the
two eclipses, and there are also signs of light curve variability in the $R$
data (which covers the largest range of time) on a timescale of
weeks. Strangely, these light curve variations seem to primarily affect the
maxima and minima.

{\bf V5 (Fig. \ref{lcwuma}):} This is a ``potential'' variable previously
identified by \citet{rh} that we confirm here (their star 170). The depths of
the eclipses in different bands differ significantly, with eclipses getting
deeper for bluer filter bands. In other close binaries, this might imply light
curve variability, yet we don't see strong evidence of this. V5 is interesting
because it is quite red compared to the main sequence, but falls very near the
base of the giant branch. However, the PLC relation implies that this is
probably a foreground system.

\begin{figure}
% clusters/n7142/ew
\plotone{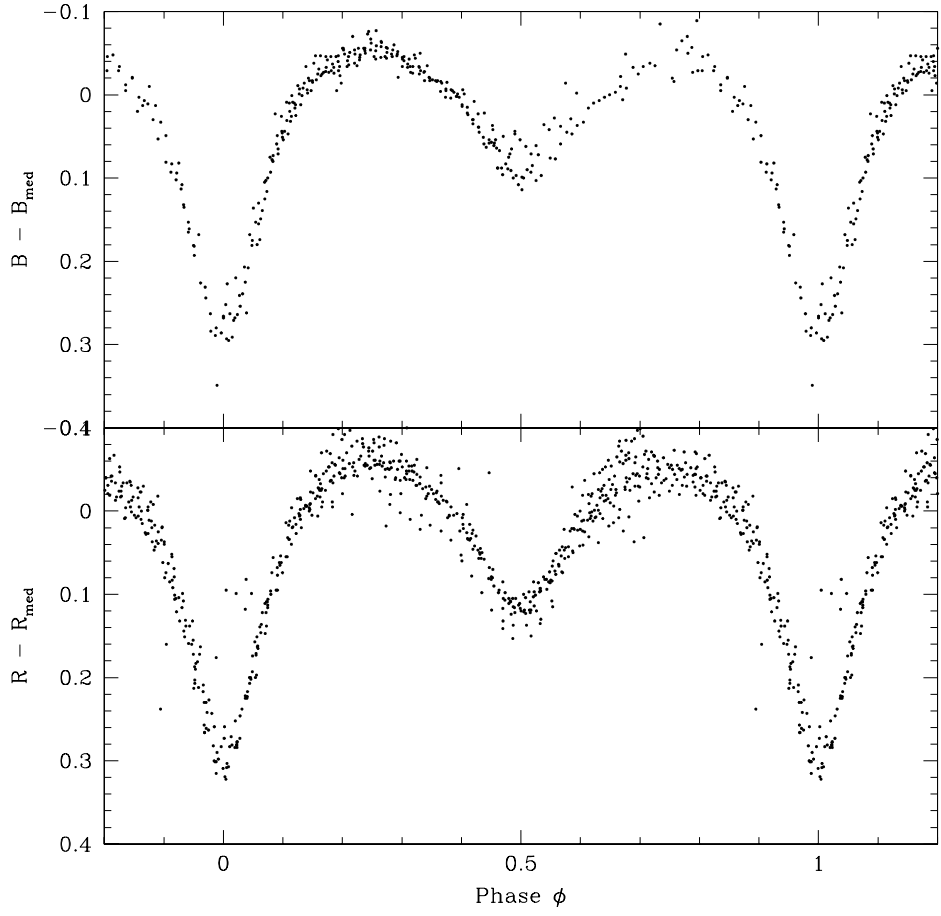}
\caption{$BR$ phased light curves for the near-contact binary V6.
\label{lcv6}}
\end{figure}

\begin{figure}
% clusters/n7142/ew/lcs.isis.v7.
\includegraphics[angle=-90,scale=0.6]{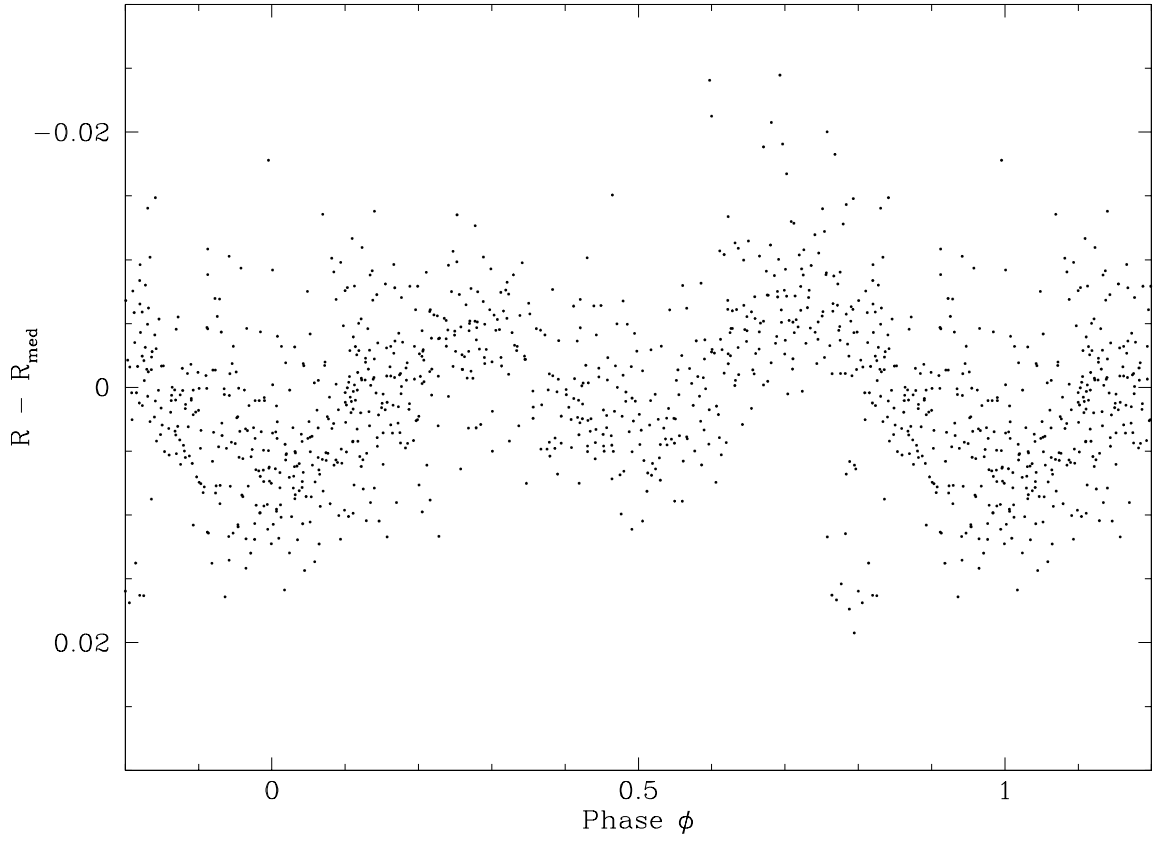}
\caption{$R$ phased light curve for the probable contact binary V7.
\label{lcv7}}
\end{figure}

{\bf V6 and V7 (Figs. \ref{lcv6} and \ref{lcv7}):} Both of these systems fall
very close to the cluster turnoff, and their variability may be related to
evolution-driven size changes if they turn out to be cluster members. The
eclipses for V6 have very different depths ($\sim 0.42$ mag versus 0.19 mag in
$R$) indicating that the stars are close, but probably not in physical
contact. As a result, the two stars may not have exchanged significant amounts
of mass and the mass of the primary star may be a future constraint on the
cluster age.

V7 is probably a W UMa-type variable with a low inclination. The light curve
shows an amplitude of about 0.02 mag as well as hints that the photometric
minimums have different depth. A period of about 0.695 d seems to be
preferred, but cycle-to-cycle variability interferes with a definitive
determination of the ephemeris.

{\bf V9 (Fig. \ref{lcwuma}):} This system has fairly deep eclipses that are
similar in depth. The combined photometry places it to the red of the main
sequence but near the expected location of the equal-mass binary sequence.

{\bf V12 (Fig. \ref{lcwuma}):} This object is toward the faint end of our
sample, and its amplitude is small and possibly variable. The system does fall
near the main sequence in the CMD though. It is difficult to find a period
under those conditions, but a period of approximately 0.29 d appears to to the
best job of phasing the observations.

{\bf V13 (Fig. \ref{lcwuma}):} V13 appears to show a total eclipse, making a
membership determination of interest. V13's corrected system photometry places
it close to the cluster main sequence, although this portion of the CMD is
fairly heavily contaminated by the field star population. The distance modulus
derived from the PLC relationship is almost a magnitude greater than
estimates for the cluster, making it unlikely to be a member. The system shows
clear variability from night to night. The most notable example occured during
a 10 day break in observations in $R$ when the system's median value dropped
by approximately 0.04 mag. During the course of the observations in $R$, the
light curve maxima varied over a range of about 0.10 mag.

{\bf V14 (Fig. \ref{lcwuma}):} This is the faintest of the near-contact
systems, and it also showed significant changes in brightness and amplitude
over the course of 2 months of observations in $R$. There appeared to be some
correlation between eclipse depth and the brightness level at the quadratures.

\subsection{Detached Eclipsing Binaries}\label{debs}

Using spectroscopic data for three of the detached eclipsing systems (V1, V2,
and V375 Cep), we simultaneously determined the center-of-mass radial velocity
and the mass ratio by $\chi^2$ minimization of a conservation of momentum
condition. For all measured radial velocity pairs, we minimized
\[ \sum (v_A - v_{CoM} + q (v_B - v_{CoM}))^2 / (\sigma_A^2 + \sigma_B^2) \]
The results are shown in Table \ref{spectab}.
%Using a map of the reduced $\chi_\nu^2 = \chi^2 / (N - 2)$, we also determined
%the uncertainties. 

There has been relatively little radial velocity work done on NGC 7142 that
can be used to judge the membership of the measured stars.  \citet{friel}
measured radial velocities for 13 stars in the field of the cluster using
relatively low-resolution spectra, finding a mean value of $-44$ \kms (12 \kms
standard deviation, and typical measurement errors of $10-15$
\kms). \citet{jaco1} measured 6 stars and found $-48.6\pm1.1$ \kms, while
\citet{spec} found $-50.3 \pm 0.3$ \kms from higher resolution spectra of 4 of
the same stars. If NGC 7142 is still in virial equilibrium, the cluster
velocity dispersion should be small ($\la 1$ \kms) given likely masses
for the cluster \citep[e.g.]{pisk}. However, the number of stars with high
precision radial velocities is still small and the distribution of field stars
has not been characterized, so membership judgements using the radial
velocities should still be made with care.

\begin{figure}
% clusters/n7142/id83
\plotone{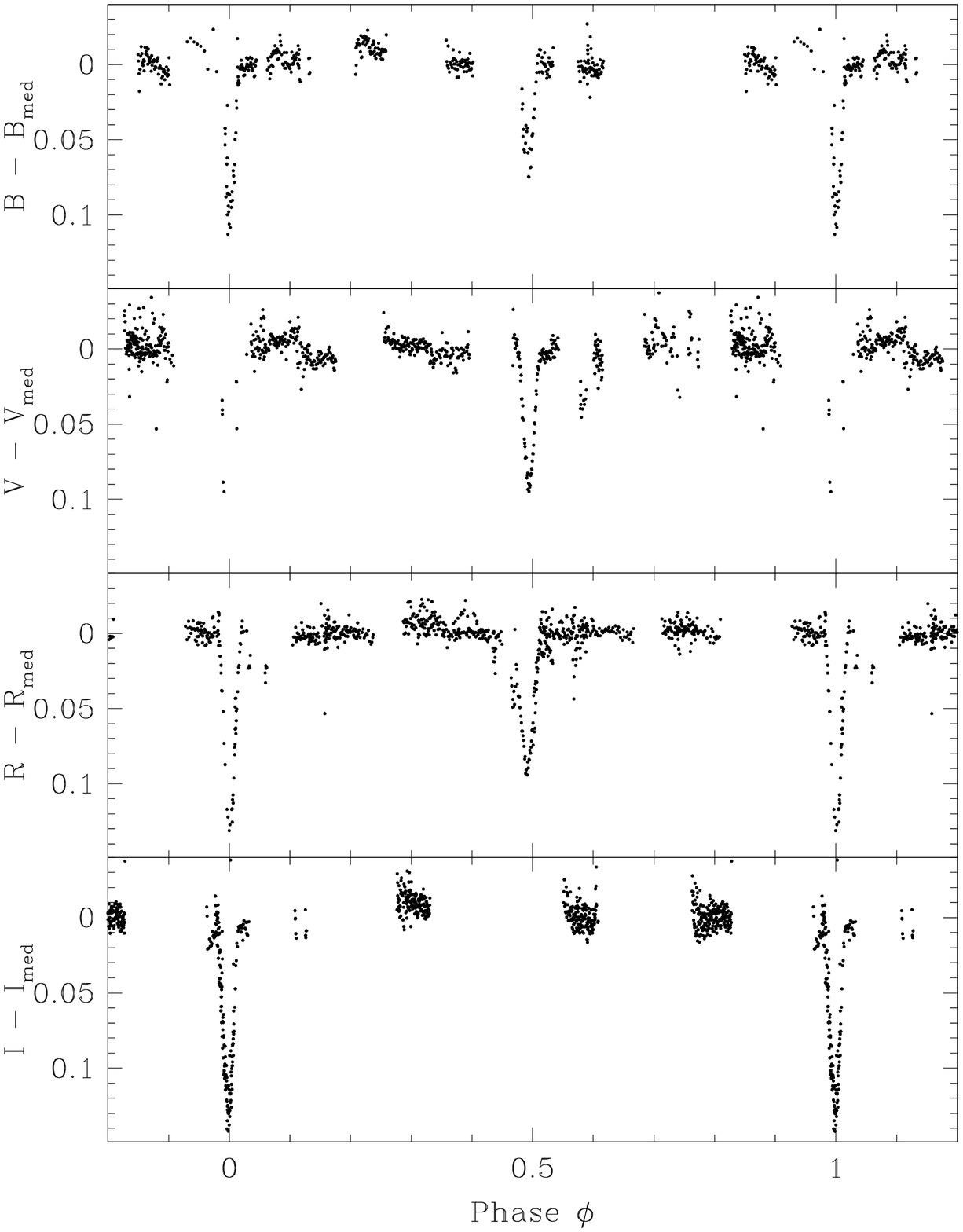}
\caption{$BVRI$ phased light curves for the detached eclipsing binary V1.
\label{lcv1}}
\end{figure}

% intrinsic noise in light curve?; fit eccen?
{\bf V1 (Fig. \ref{lcv1}):} Primary and secondary eclipses were initially
detected with comparable depths, making this very likely to be a double-lined
spectroscopic binary with stars of comparable mass. We subsequently confirmed
two detectable components in spectra taken with the HET. Although the system
falls in the blue straggler portion of the color-magnitude diagram, there was
a possibility that the components are main sequence stars because differential
reddening may affect the position of the blend in the CMD. Our three radial
velocities imply a mass ratio $q = 0.96 \pm 0.02$ and a system velocity $v_r =
-17.0 \pm 1.0$ \kms. The system velocity puts it far
away from the likely cluster mean.  Based on the smalll displacement of
the secondary eclipse from phase 0.5, the binary also has at least a small
eccentricity. Binaries with periods this short ($< 5$ d) appear to be able to
circularize on timescales much shorter than NGC 7142's age \citep{mandm},
although on its own this is not conclusive because some known blue stragglers
have short periods and significant eccentricity. But based on these indications,
we conclude it is most likely not a member of the cluster.

% both components have smaller eclipse depths in bluer filters: hot third
% component (outside of binary)?

\begin{figure}
% clusters/n7142/v2
\plotone{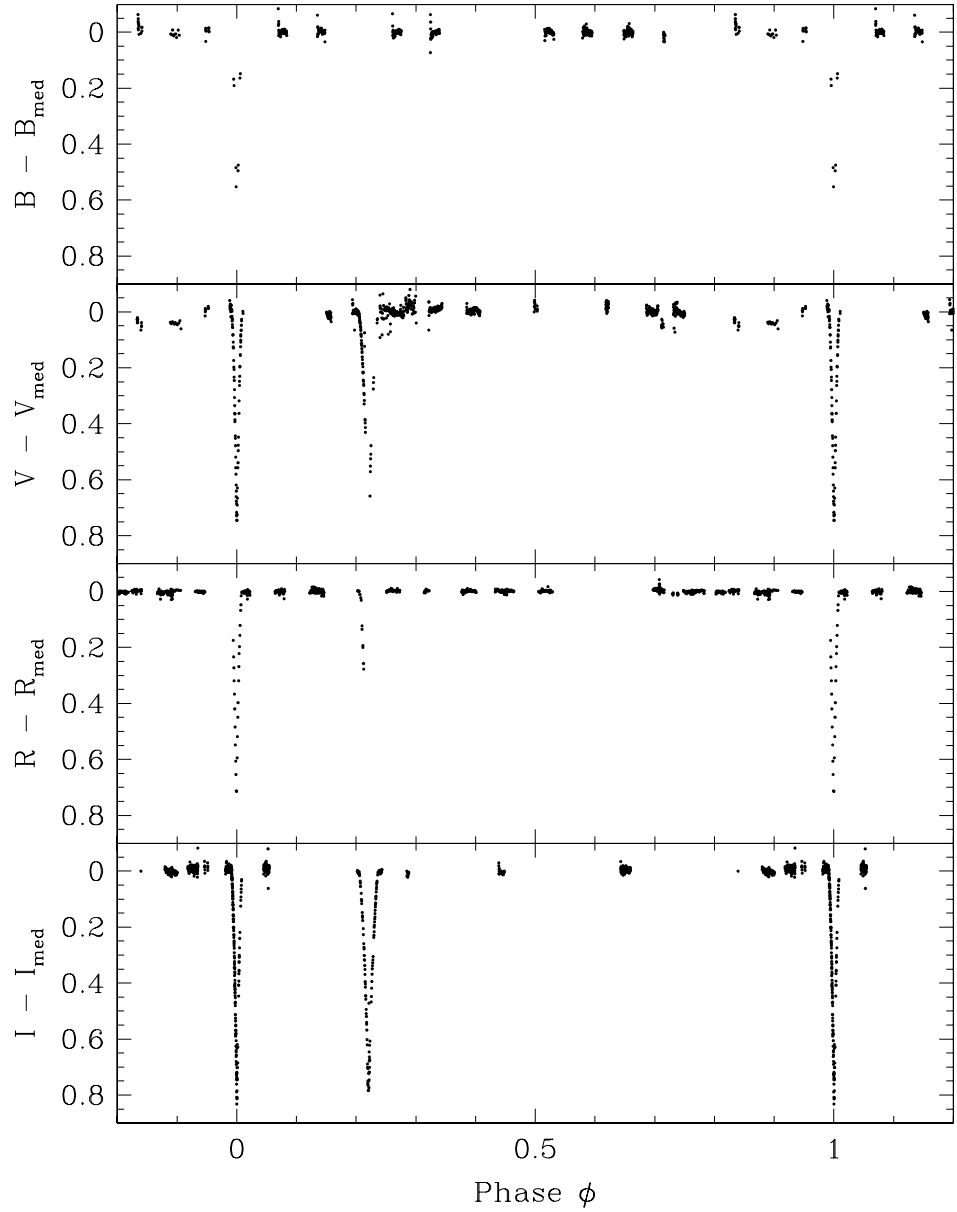}
\caption{$BVRI$ phased light curves for the detached eclipsing binary V2 near
  its eclipses. Note the longer duration of the secondary eclipse, and its
  displacement from a phase of 0.5.
\label{v2}}
\end{figure}

{\bf V2 (Fig. \ref{v2}):} This appears to be a rather extraordinary eclipsing
binary system. It was first identified in a single deep ($\sim 0.75$ mag)
eclipse in our observations from 2005. While characterizing the light curve of
V375 Cep (see below) in 2008, we detected two additional eclipses (a primary
and a secondary) separated by about 19.03 d. At that time, we noticed that the
secondary eclipse ingress was significantly longer in duration than the
primary eclipse, implying a significant eccentricity. However, the large
separation between the first and second observed eclipses (1017.3 d) in
combination with the eccentricity made it difficult to determine the period.
Using the photometric constraints along with three subsequent radial velocity
measurements, we were able to narrow down the possibilities by requiring i) no
photometric observations at constant light overlapped with predicted eclipses,
and ii) radial velocities should have a physically realistic distribution in
phase (e.g., no sudden changes in direction implied when phased). Using these
constraints, the system was observed on dates with eclipses predicted for the
most likely periods until a period of 15.6505 days was finally confirmed.

With this period, the secondary eclipse is centered near phase $\phi = 0.22$,
consistent with a large eccentricity. 
%The phase position and shape of this eclipse strongly
%constrain the longitude of periastron $\omega$.  
The spectra of the system to date confirm that it is double-lined and that the
system velocity ($-42.1\pm0.6$ \kms) falls near the mean cluster value, but
different by about several \kms. At this time, we have to regard V2 as a
possible cluster member. In spite of the measured mass ratio ($q =
0.99\pm0.10$), the two stars differ significantly in size, showing the effects
of evolution. When the depths of the eclipses are considered, the implied
inclination of the orbit is very close to $90\degr$.

In the CMD, V2 falls in the blue straggler region, brighter and bluer than the
turnoff. Both of the stars individually have magnitudes placing them near the
turnoff, and one or both may still be bluer than the turnoff when their light
is disentangled. At this time, however, we cannot rule out the possibility
that they are on a line of sight having less than the average cluster
reddening.  The characteristics of the orbits in V2 don't appear to
distinguish between these possibilities: the period at which most binaries
have circularized in M67 (which has similar age to NGC 7142) is approximately
$12.1^{+1.0}_{-1.5}$ d \citep{mandm}.

% E(B-V)=0.41
% E(V-K)/E(B-V) = 2.63 Cardelli et al. 1989, E(V-K) = 1.0783
% system (unreddened) V=15.310+/-0.009 B-V=0.797+/-0.012 V-I=1.010+/-0.013; 
%   2MASS: 13.661+/-0.029 H=13.342+/-0.032  K=13.227+/-0.036  V-K=2.083
%  (V-K)0 = 1.005
% DiBennedetto et al : log T = 3.816   T=6550 K
% Casagrande et al. 2010  (V-K): 6690 K   (B-V): 6725 K  (V-I)_C: 6604 K
% spectroscopic star CT387 has V=13.378  B-V=0.776 V-I=0.923

\begin{figure}
% clusters/n7142/v375
\plotone{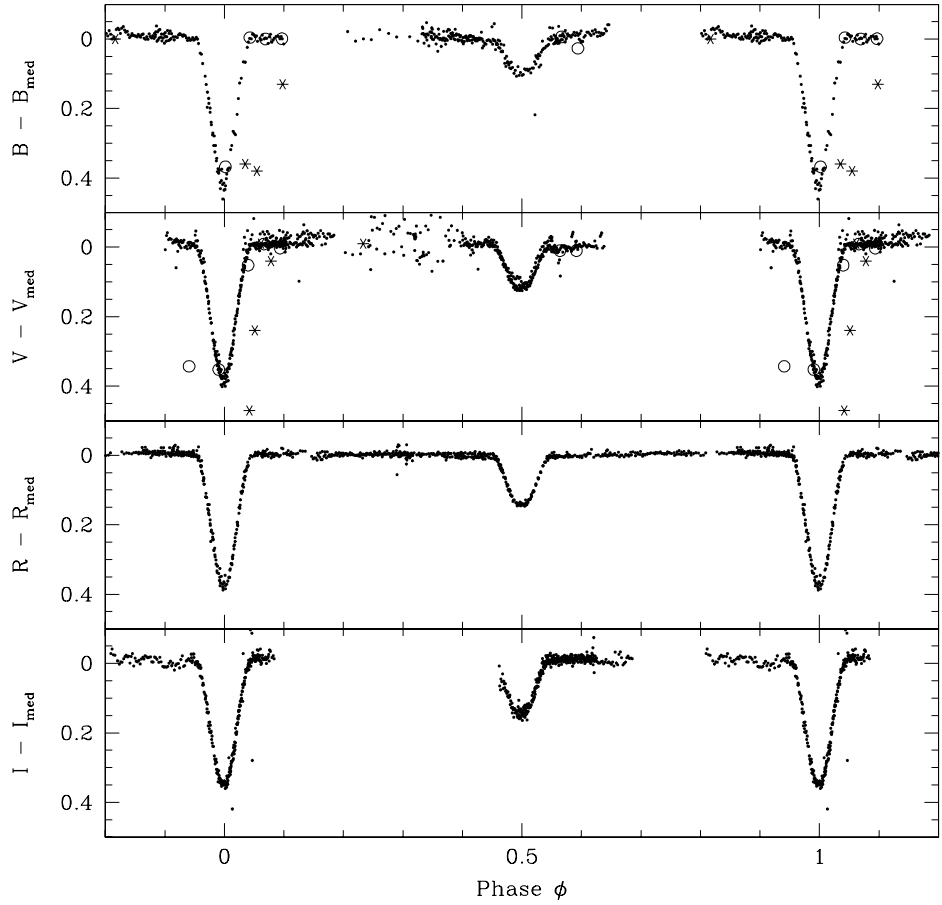}
\caption{$BVRI$ phased light curves for the detached eclipsing binary V375
  Cep. Open circles indicate measurement made by \citet{ct} and asterisks are
  measurements made by \citet{see} phased to our ephemeris.
\label{v375}}
\end{figure}

{\bf V375 Cep (Fig. \ref{v375}):} This system was the primary motivation for
many of the observations we took, and so we have observations of primary and
secondary eclipses in all photometric bands. The secondary eclipse is clearly
detected, and we have shown that it is a double-lined spectroscopic binary.
The spectroscopic data give $v_{CoM} = -49.1 \pm 1.7$ \kms and $q = 0.69 \pm
0.03$, and a minimum $\chi_\nu^2 = 0.98$. $v_{CoM}$ is completely consistent
with the cluster average (see the beginning of this section), making it a
highly probable cluster member.

We collected a small number of previous photometric observations from
\citet{ct} and \citet{see} to improve the accuracy of the ephemeris and test
for the possibility of a nonlinear ephemeris using the 27 y baseline. 
For the densely observed light curves from our study, we used the method of
\citet{kwee} to determine times of minima and the errors.

Most of the observations by \citet{ct} agree well with our phased light curve
and confirm that they observed the system very near one of the primary
minima. They had observations in and out of eclipse on the night of one
eclipse, which allowed us to fit our $BV$ light curves to their data and
derive an approximate time of minimum. One additional observation in $V$ on a
different date also appears to have fallen near an eclipse minimum. If there
are no errors in their Table 2, the separation between their two faintest
measurements in $V$ (26.83 d) would be consistent with a period of about
1.91647 d. However, the implied change in period over the 20 years since the
\citeauthor{ct} observations seems unrealistically large. A linear ephemeris
that satisfactorily matches our data and one of the \citeauthor{ct} eclipse
observations would cause a disagreement of about 0.06 in phase for the other,
which is slightly longer than the entire duration of an eclipse ingress.

In addition, most of the \citeauthor{see} eclipse measurements disagree
significantly with a linear ephemeris based on our data. They appear to have
caught three measurements during the egress of a primary eclipse on one night
(HJD 2446650), although the steeper slope and larger depth of their eclipse of
when compared to ours makes it hard to trust estimations of the time of
minimum. We have therefore elected to assign asymmetric error bars to that
point. Two faint measurements in $B$ seem likely to have been taken during
ingress and egress respectively, if the period has remained near 1.9
d. However, another observation in $B$ later on one of those nights was not
consistent with out-of-eclipse levels or with predictions based on our eclipse
light curves. Unfortunately it appears that the early data on this variable is
not of sufficient quality to test for nonlinearities in the ephemeris.

\begin{figure}
% clusters/n7142/id504
\plotone{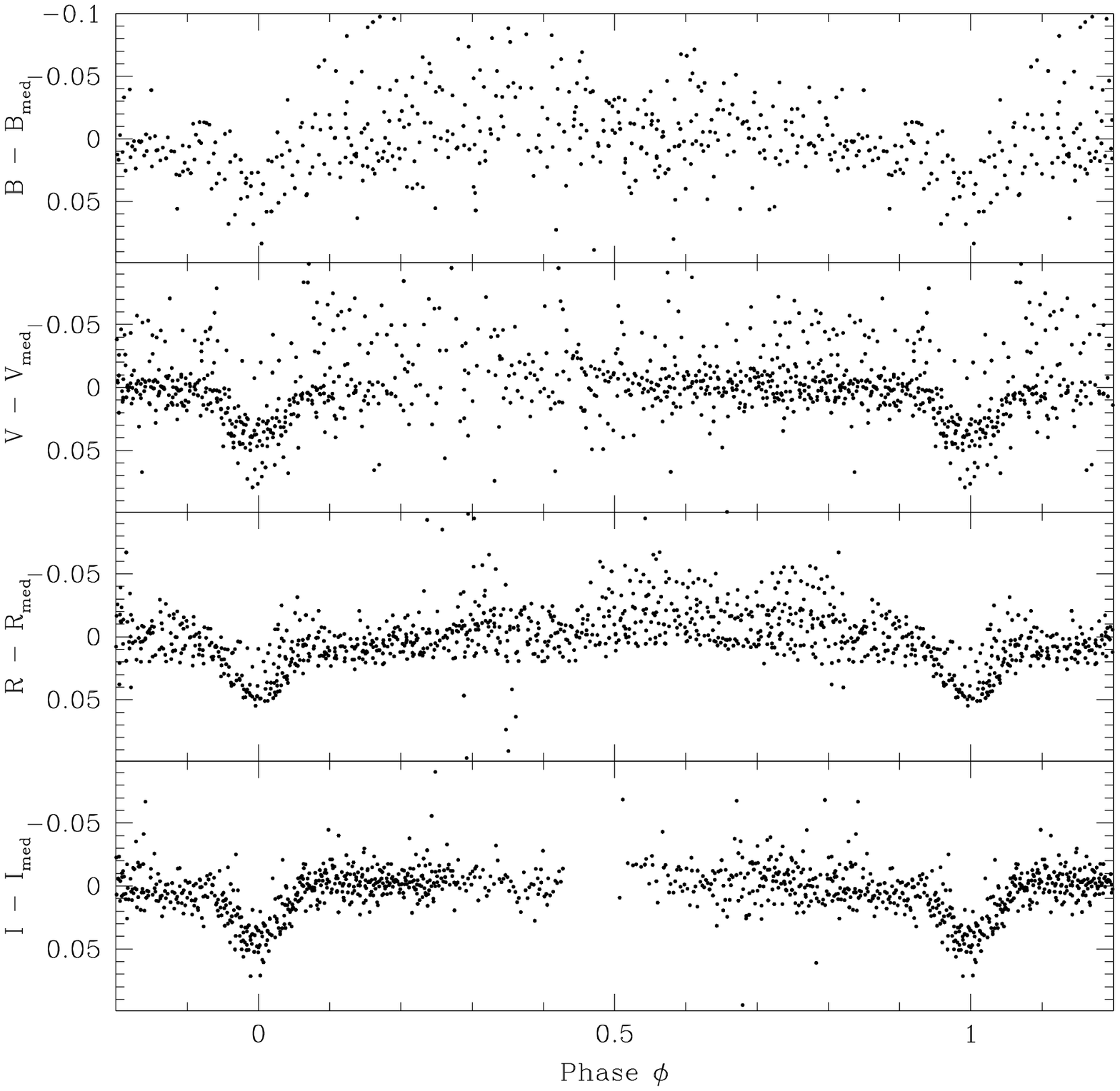}
\caption{$BVRI$ phased light curves for the detached eclipsing binary V8.
\label{lcv8}}
\end{figure}

{\bf V8 (Fig. \ref{lcv8}):} This is a binary with partial eclipses of quite
small amplitude ($\sim 0.03-0.04$ mag). Eclipses were detected on some
consecutive nights separated by about 1.0546 days, although we find the light
curves phase together better when half this period is used. If the true period
is indeed near 0.53 d, a secondary eclipse is not detected.

This system falls near the expected position of the equal-mass binary
sequence, so it has a reasonable probability of cluster membership. However,
if it is a cluster member, it would need to be a triple system, as this would
explain both the shallow eclipses (through the diluting effects of third
light) and the position in the CMD brighter than the main sequence. 
A third star with brightness similar to the
primary of the binary system would need to be on a wider orbit. If the system
is not a cluster member, it could involve a star being eclipsed by a smaller
faint object. The significant out-of-eclipse variation implies that the
companion must be of stellar mass in order to significantly distort the
primary.

% produce averaged curves?

\begin{figure}
% clusters/n7142/id689
\plotone{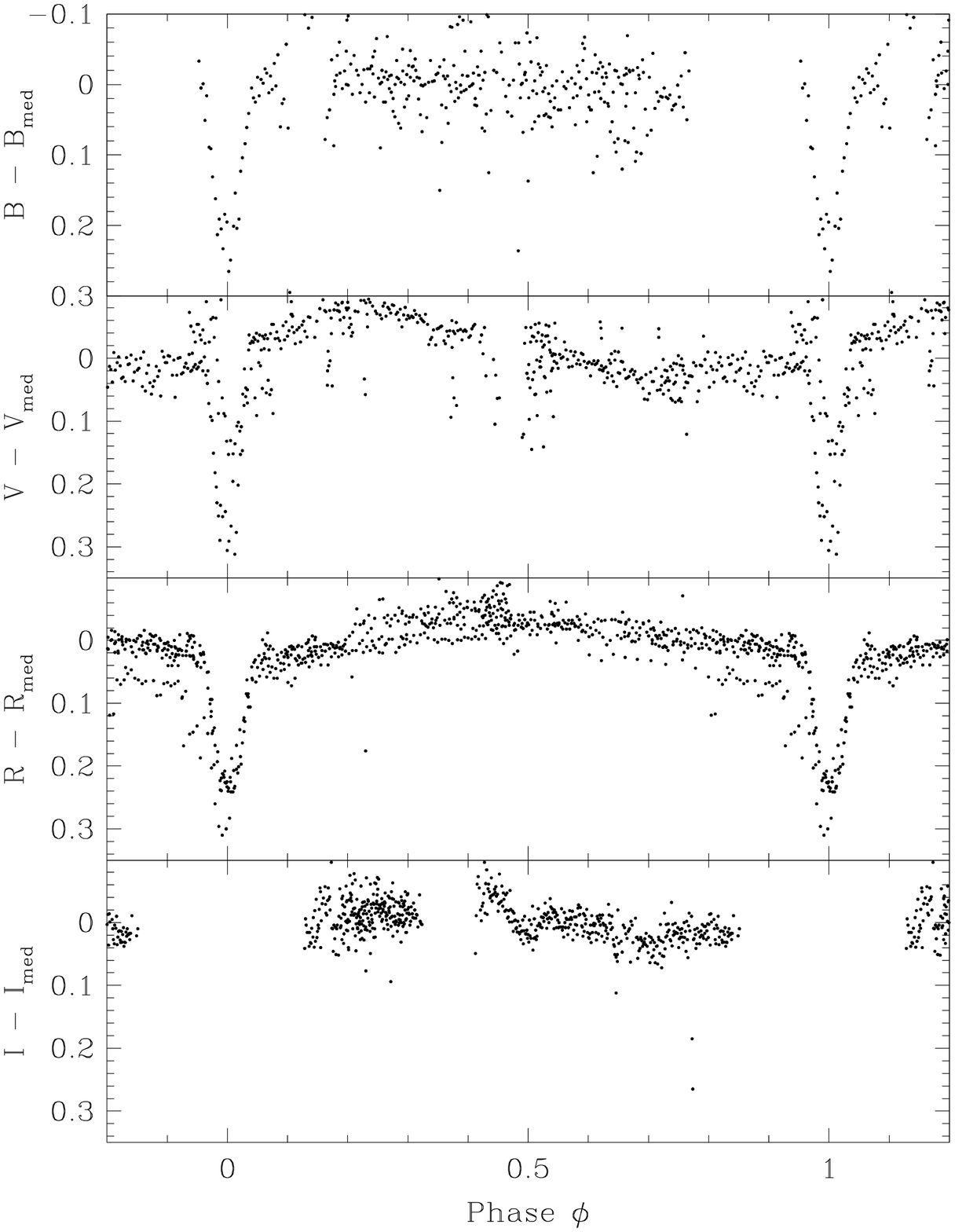}
\caption{$BVRI$ phased light curves for the detached eclipsing binary V11.
\label{lcv11}}
\end{figure}

{\bf V11 (Fig. \ref{lcv11}):} Eclipses were detected in $R$ at HJD 2453594.97,
2453598.83, 2453638.70, and 2453639.99, in $V$ at 2454676.74, 2454678.02 (only
ingress), and 2455081.93, and in $B$ at 2454627.85. In two cases, eclipses of
nearly equal depth ($\sim 0.2$ mag) were observed on consecutive nights, and
in one additional case, eclipses were separated by about 3.85 d. These
observations make the period very likely to be near 1.3 d. Phasing the
observations to this period, it appears we may have detected a shallow
secondary eclipse in $I$ band only, centered at phase 0.5 with approximate depth
0.03 mag. This system is therefore likely to be a single-line spectroscopic
binary with a fairly small mass ratio.  Although we do not have radial
velocity measurements for this system, the position of the system in the CMD
implies that there is a good chance of cluster membership.

The primary eclipses appear to have a short period of totality. We also see
strong out-of-eclipse light curve variations. For example, observations in
2005 separated by a little more than a month show large differences in the
out-of-eclipse phases. In addition, a later night of observations in $R$
shows the system brightness increasing before an eclipse, contrary to other
observations at similar phase. The variations may be due to strong spot
activity driven by rapid rotation in a system with such a short orbital
period.
% drops seen frequently (Esp. in B) mostly garbage, not eclipse related

\subsubsection{Other Variables}

%ID 82 has small amplitude ($\sim 0.02$ mag) variability and multiperiodic
%short timescale variability consistent with being a $\delta$ Scuti star. 
%Its CMD position (redder than the giant branch) is inconsistent with being a
%$\delta$ Scuti star however.

\begin{figure}
% clusters/n7142/irr/
%\epsscale{0.3}
%\plotone{lcavv4.ps}
%\includegraphics[angle=90]{lcavv4.ps}
\includegraphics[scale=0.5,angle=270]{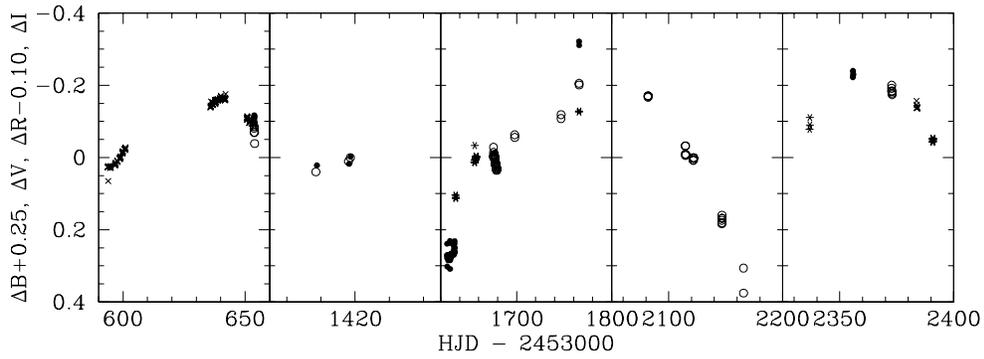}
\caption{Light curves (magnitude minus the median value in each filter band)
  for the long-period variable V4 in $B$ ($\bullet$), $V$ ($\bigcirc$), $R_C$
  ($\times$), and $I_C$ ($\ast$). Individual points are averages of
  measurements made during time intervals up to 0.05 d. Measurements in $B$
  and $R_C$ have been shifted to bring them into approximate agreement with
  $V$ measurements on HJDs 2454419 and 2453653, respectively.
\label{lcv4}}
%\plotone{lcv4.ps}
%\caption{Light curves for the long-period variable V4. Individual points are
%  averages of measurements made during time intervals up to 0.05 d. Error bars
%  are also plotted, but are in most cases indistinguishable from the point.
%\label{lcv4}}
\end{figure}

V4 shows variability on multiple timescales and with different amplitude.  We
have averaged measurements for each night in Fig. \ref{lcv4} to make these
trends clearer. Because of the large numbers of observations, the errors in
most of these averages are smaller than the sizes of the points.  During
individual nights of observation there are small but significant trends of
about 0.01 mag, particularly noted in our $B$ and $V$ observations. On the
other hand, our observations reveal longer timescale variations (tens of days)
of greater amplitude (at least 0.6 mag in $B$, $V$, and $R$). Comparisons of
observations of measurements in different filter bands should be regarded as
approximate --- the medians in each band generally do not correspond to the
identical times in the oscillation cycle, and there is evidence that the
variation amplitude varies from band to band.

V4 appears to be an single star in our best seeing images. The CMD position of
the star sets it apart --- it is quite red even for a star on the giant
branch, and even accounting for realistic amounts of differential
reddening. On the night of our calibration observations (HJD 2454765), the
star appears to have been in a bright phase: more than 0.6 mag above the
median of our $B$ measurements, and about 0.2 mag above the median in
$V$. (Note that we generally did not take observations in multiple bands
during one night, so the median measurements are probably not representative
of corresponding points in the variability cycle of the star.)  However, V4
resides in a similar position in the CMDs of \citet[][star Y]{vh} and
\citet{ct}. Our calibration observations and the observations of these other
photometric studies were taken close together in time, so that they should be
fairly representative of the object's instantaneous color. Because they place
the star to the red of the cluster giant branch, the calibrated color cannot
be representative of a static star if it is a cluster member as it would be
redder than the Hayashi line.

The nature of the variability and its CMD position are somewhat reminiscent of
two variable members of the similarly old cluster M67
\citep{vdbm67,ssm67}. S1063 and S1113 sit below the subgiant branch of M67
\citep{subsub} and are both X-ray sources and spectroscopic binaries. S1113
shows fairly periodic variations, while S1063 does not. In both cases, the
variations are of much smaller amplitude than we see in V4. We have not
detected variability among stars below the subgiant branch in NGC 7142, and
they would be impossible to identify based on CMD position alone, thanks to
field star contamination. We also carefully examined the other object that
resides near V4 in the CMD, but saw no sign of variability. Spectroscopic
information would be valuable to verify the stellar nature of the star, to
check cluster membership (both by measuring radial velocities and by checking
the possibility that it is a background giant).

\begin{figure}
% clusters/n7142/irr/
%\epsscale{1.0}
\plotone{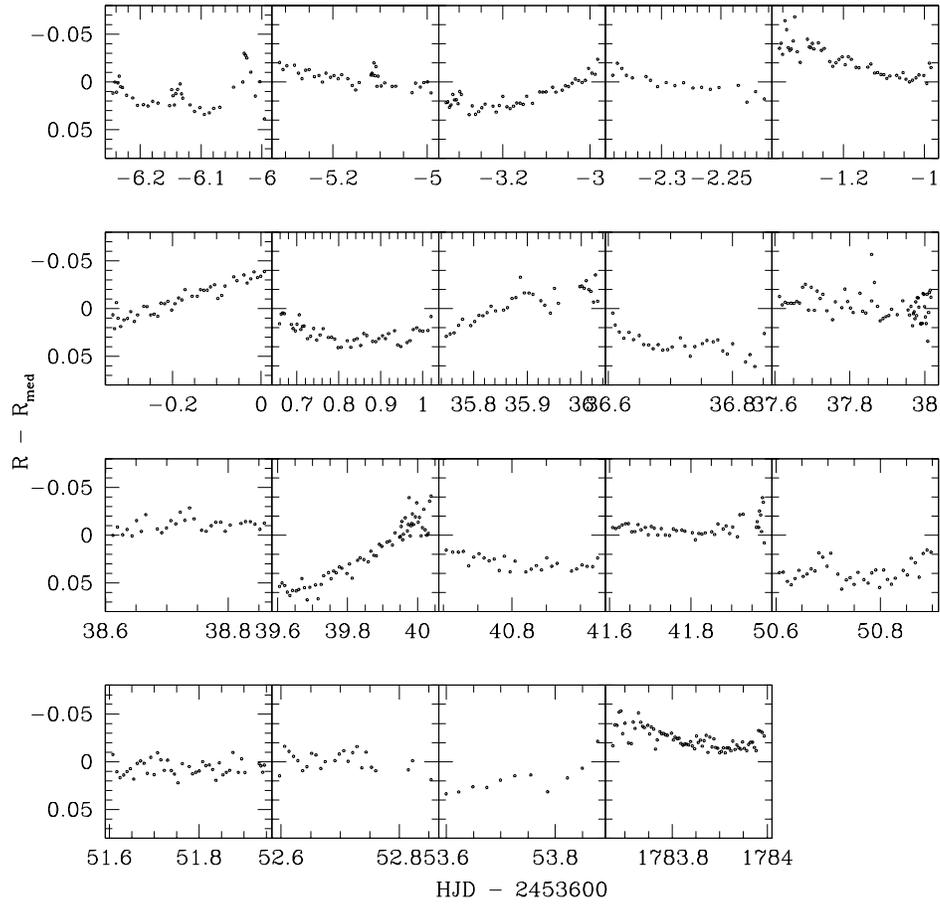}
\caption{$R$ light curves for the quasiperiodic variable V10.
\label{lcv10}}
\end{figure}

V10 appears to be a quasiperiodic variable. On any given night the star shows
a trend in brightness of up to about 0.1 mag from beginning to
end. Fig. \ref{lcv10} shows our observations in $R_C$, but similar variation
appears in other filters as well. Using the Lafler-Kinman method \citep{lk},
we found the most likely periods of variability to be approximately 1.38 and
2.77 d, but we have not been able to find a period that phases all of the
datasets satisfactorily. The object is well separated from other sources, so
that contamination of the photometry is unlikely. Based on its CMD position
redder than the equal-mass binary sequence, it is not likely to be a cluster
member. However, its unusual variation may be worth further study.

\section{Discussion}

Overall, little work has been done on NGC 7142 compared to other open
clusters, but as will be seen below, there is reason to believe the cluster's
characteristics should be revised. Before addressing the age of the cluster
with a traditional isochrone fit, we first review previous estimates of the
metallicity, distance, and reddening.

\subsection{Metallicity}

In the last decade or so, spectroscopic measurements of NGC 7142's metallicity
have been moving more metal-rich. This trend of cluster metallicities moving
more metal-rich with time is not unique to this cluster (see \citealt{gratton}
for NGC 6791, for example), but the fact that the cluster now seems to be
slightly more metal-rich than the Sun is important for accurate age
determination. The first spectroscopic measurement for NGC 7142 by \citet{fj}
([Fe/H] $=-0.23\pm0.13$) as part of a larger survey of open clusters was later
revised upward to $-0.10\pm0.10$ by \citet{friel2002} based on a new
calibration of Fe spectroscopic indices.  \citet{twa} computed an even higher
value ($+0.04\pm0.06$) by transforming the same \citet{fj} data to a
common metallicity scale based on DDO photometry. Later spectroscopic
measurements ($+0.08\pm0.06$, \citealt{jaco1}; $+0.14 \pm 0.01$ from 4 giants,
\citealt{spec}) also indicated a super-solar metallicity, however. NGC 7142
also appears to be more metal rich than the well-studied cluster M67 ([Fe/H]
$=0.00 \pm 0.09$, \citet{twa}; $+0.03 \pm 0.07$, \citet{friel10}) in studies
where the two clusters can be compared reliably.

\subsection{Distance and Reddening}\label{dist}

\citet{vh} determined that the reddening for the cluster was $E(B-V) = 0.41$
using the $UBV$ color-color diagram, and this has been the most commonly
quoted value since. \citet{ct} found evidence of differential reddening on the
order of $\Delta E(B-V) = 0.1$ based on the width of the observed main
sequence. Even though there are clear signs of a cloud front stretching ENE to
WSW (with the amount of gas appearing to decrease toward the SSE),
\citeauthor{ct} did not find evidence of a systematic change in reddening
across the face of the cluster. They interpreted this to mean that there is
significant smaller scale variation. \citet{twa} settled on $E(B-V) = 0.43$
for turnoff stars and $0.40$ for giants, based on a survey of previous studies
including the two above.  \citet{janes} derived $E(B-V) = 0.32 \pm 0.05$ using
a method depending largely on comparing photometry of the red giant clump and
main sequence turnoff to those of synthetic CMDs generated from isochrones.
The reddening derived from IRAS and COBE dust maps \citep{sfb} is
significantly larger [$E(B-V) = 0.51\pm0.02$] than any previous determination
for the cluster, but this might be a better indication of the reddening of
stars beyond NGC 7142. The dust map values do vary in a range covering 0.09 mag
within $5\arcmin$ of the cluster center\footnotetext{\tt
  http://irsa.ipac.caltech.edu/applications/DUST/}, lending some credence to
the possibility of differential reddening for the cluster.
% possibility of overestimation in absolute value, but less likely in variation

There has not been much agreement among previous distance modulus
determinations either. The first estimate by \citet{vh} was $(m-M)_V \simeq
13.7$. \citet{ct} quoted a dereddened distance modulus $(m-M)_0 = 11.4\pm 0.9$
when using $E(B-V) = 0.41$, corresponding to an uncorrected distance modulus
of approximately $(m-M)_V = 12.67$. \citet{twa} derive $(m-M)_V = 12.95$ from
main sequence fitting, assuming $E(B-V) = 0.43$. \citet{janes} derive $(m-M)_0
= 11.85\pm0.05$, or $(m-M)_V = 12.84$.

\begin{figure}
% clusters/n7142/2mass/cmd.optir.
\epsscale{0.7}
\plotone{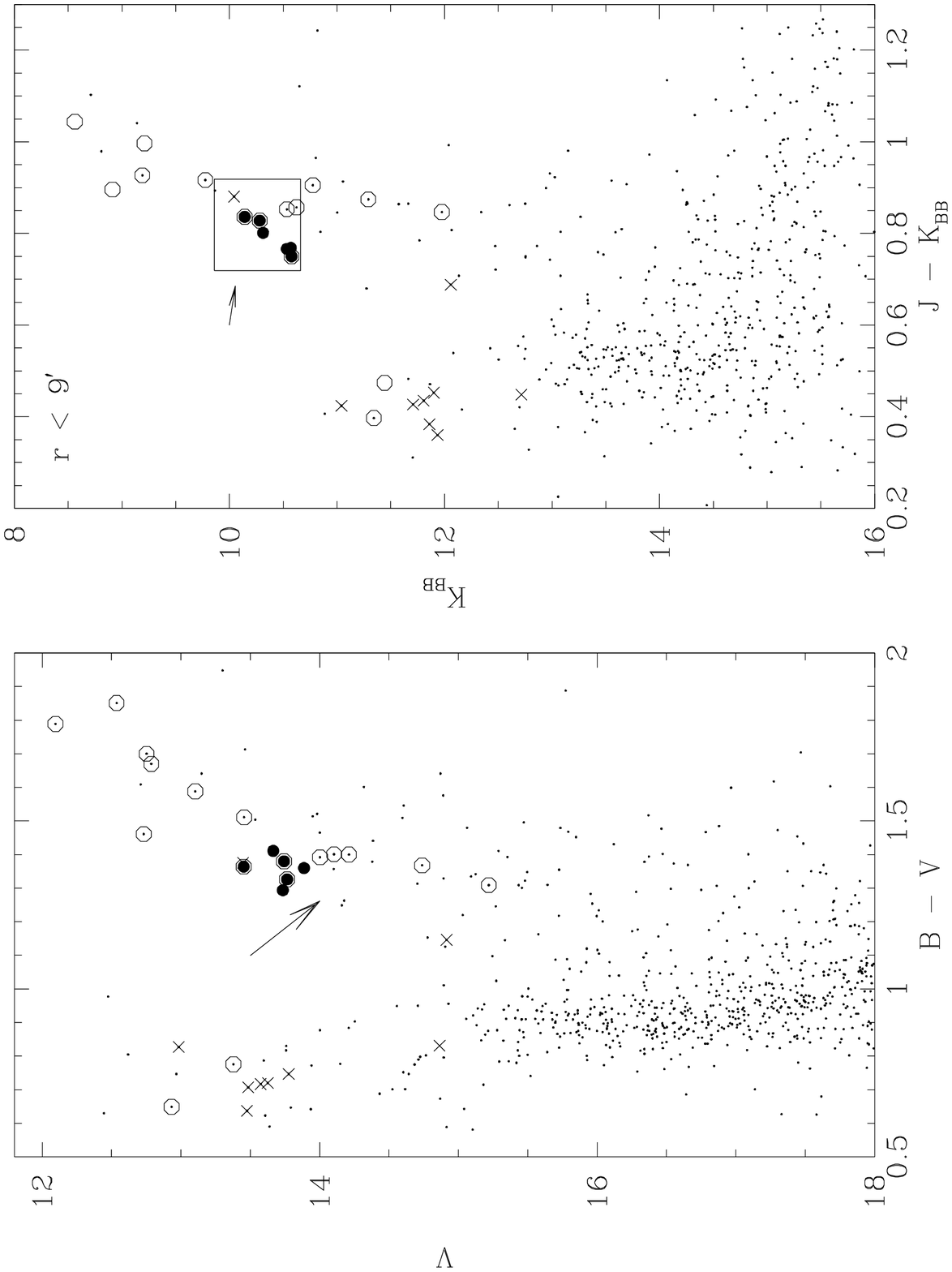}
\caption{Optical (from the current study) and infrared (2MASS) CMDs for NGC
  7142. Red clump candidates are identified with $\bullet$, likely spectroscopic
  cluster members are shown with $\bigcirc$, and likely nonmembers are shown
  with $\times$. The box shows a \citet{groc} red clump selection box shifted
  according to \citet{janes} values for reddening and distance
  modulus. Reddening vectors are also shown, with lengths for both
  corresponding to $A_V = 0.5$ mag.
\label{clumpcmd}}
\end{figure}

Given the disagreements, it is worth revisiting the issue here.  Of the
features in the CMD on NGC 7142, the red giant clump is probably the most
clearly identifiable, and some stars within it have been identified
spectroscopically as likely cluster members. We identified candidate clump
stars in optical and 2MASS infrared CMDs (see Fig. \ref{clumpcmd}), and the 6
stars are given in Table \ref{rctab}. Obviously, the sample is small, and
there may be a field star or first-ascent giant star present, given the amount
of differential reddening across the cluster's face. In fact, one star (CT
399) was rejected based on radial velocity measurement \citep{jaco1}. We
obtained single radial velocity measurements for three other red clump stars
(listed in Table \ref{spectab}), and one of them (JH 2222) is consistent with
membership, while the other two have velocities that are separated from the
cluster mean by $\sim4$ \kms and whose memberships are still in
question. However, by using the median magnitude and color of the sample in
2MASS photometry (as did \citealt{groc} in their study of open cluster clump
stars), the problems related to membership determination should be minimized.

%However, we find that there are two
%separate groupings of stars that could be considered clump stars. In their
%optical study, \citet{janes} identified the clump at $V = 13.76$, $(B-V) =
%1.36$, and $(V-I) = 1.50$, and 4-5 of the stars we identified as clump star
%candidates are consistent with that. In a 2MASS CMD, however, only two of
%these stars remain in a bright group ($K_{BB} \approx 10.2$), and a group of 6
%stars can be identified at $K_{BB} \approx 10.5$. Most of this fainter group
%appears at $V \approx 14.05$ in the optical.  As one can see, the
%identification of the clump is not trivial, and it has a large ($\sim 0.3$
%mag) effect on the distance.

%            N7142 med    M67 med     DA             AK=0.11AV --> AV=0.93 --> AK=0.10 --> DM=2.32
% A_B         15.08        11.63      3.45    AB/AV=1.215
% A_V         13.74        10.49      3.25
% A_I         12.27        9.43       2.64    AI/AV=0.344
% A_J         11.13        8.59       2.54
% A_H         10.532       8.10       2.43               (10.263,10.399,10.698,10.438,10.626,10.710)
% A_KS        10.377       7.96       2.42
%--> higher implied R_V?

The color of the clump is largely a function of age in the range of
composition containing NGC 7142 [$\partial(J-K_{BB})_0/\partial\log t_{age}
\approx -0.17 $ dex$^{-1}$] according to the open cluster data of
\citet{groc}.  Based on their study, the clump is predicted to have a color
$(J-K_{BB})_0 = 0.67\pm0.03$, where we have assumed an age similar to M67 (4
Gyr) with a conservative uncertainty of about 1.5 Gyr. For NGC 7142, we
find a median observed color $(J-K_{BB}) = 0.80$, for a reddening $E(J-K_{BB})
= 0.13\pm0.03$ or $E(B-V) = 0.25\pm0.06$.
% if (J-K)_0 = 0.66, E(J-K)=0.17, E(B-V) = 0.34

Using the median magnitude for NGC 7142 stars ($K_{BB} = 10.42$), and the
median absolute value from \citeauthor{groc} for stars similar in age and
metallicity to NGC 7142 ($\langle M_K \rangle = -1.62 \pm 0.06$), we find
$(m-M)_K = 12.04 \pm 0.15$.  These values imply $(m-M)_0 = 11.96$ or $(m-M)_V
= 12.72\pm0.19$. The reader should note that the 6 most probable clump stars
are separated into two groups in $K$ magnitude and $(J-K)$ color, and the
medians split the difference here. In the optical CMD used by \citet{janes},
there is a greater degree of consistency in the $V$ magnitudes of the clump
stars, which is somewhat odd considering the apparent differential reddening
across the cluster.

After comparing with the results of previous studies, the two results
derived from the clump (this study and that of \citeauthor{janes}) give the
largest values for the distance and smallest values for the reddening (with
the exception of the early estimates by \citealt{vh}). Earlier estimates were
probably affected by degeneracy between the effect of dust extinction and
distance, particularly because other features (like the turnoff) are difficult
to identify in the cluster's CMDs.  The larger distance implies that the
turnoff is more luminous than previously thought and that the age will tend to
be on the low side, as discussed below.

% file in n7142cal.clump.dat

% younger age implies bluer clump, giving larger reddening
% E(J-K) / E(B-V) = 0.53; E(J-K) = 0.17 AV; 
% 0.1 error in log age--> 0.017 error in color --> 0.032 error in E(B-V); 0.1 mag error in AV

\subsection{Age}\label{aged}

Part of the motivation for this project involved finding detached eclipsing
binary star systems that contained evolved stars. Precise measurements of the
masses and radii for stars that have evolved off the main sequence can lead to
precise age determinations for the cluster. A detailed analysis of the two
detached eclipsing binary systems sitting at the cluster turnoff (V375 Cep and
V2) and the implications for the cluster age will be presented in an upcoming
paper. However, it is beneficial to have as reliable an age determination from
isochrones and the CMD for comparison.

With the current CMDs, it is not easy to tell whether NGC 7142 turnoff stars
are high enough in mass to produce a significant ``hook'' feature associated
with core convection on the main sequence. A detailed isochrone fit for a
feature of that kind is complicated by the short evolution timescale for stars
moving through the hook, and the difficulties are exacerbated by differential
reddening across the face of the cluster, field star contamination, and
confusion due to optical blends of stars or true binaries.  There is a weak
indication of a subgiant branch at $V \approx 15.8$, sloping faintward from
the massive end of the main sequence, but more work is needed to establish the
cluster membership of these stars.

As a first check, we compare with M67, a well-studied cluster with an age
thought to be near that of NGC 7142 \citep{vh,ct}. M67 has a low and
well-determined reddening ($E(B-V)=0.041\pm0.004$; \citealt{taylor}), and a
well-determined distance modulus [$(m-M)_V = 9.72 \pm 0.05$, $(m-M)_0 = 9.60
  \pm 0.03$; \citealt{sandm67}]. M67 also appears to be a transition object,
with evidence of weak core convection for the most massive turnoff stars, and
a sparsely populated subgiant branch. 

\begin{figure}
% clusters/n7142/isos/cmdvsm67.2.mc.
\epsscale{0.8}
\includegraphics[scale=0.5,angle=270]{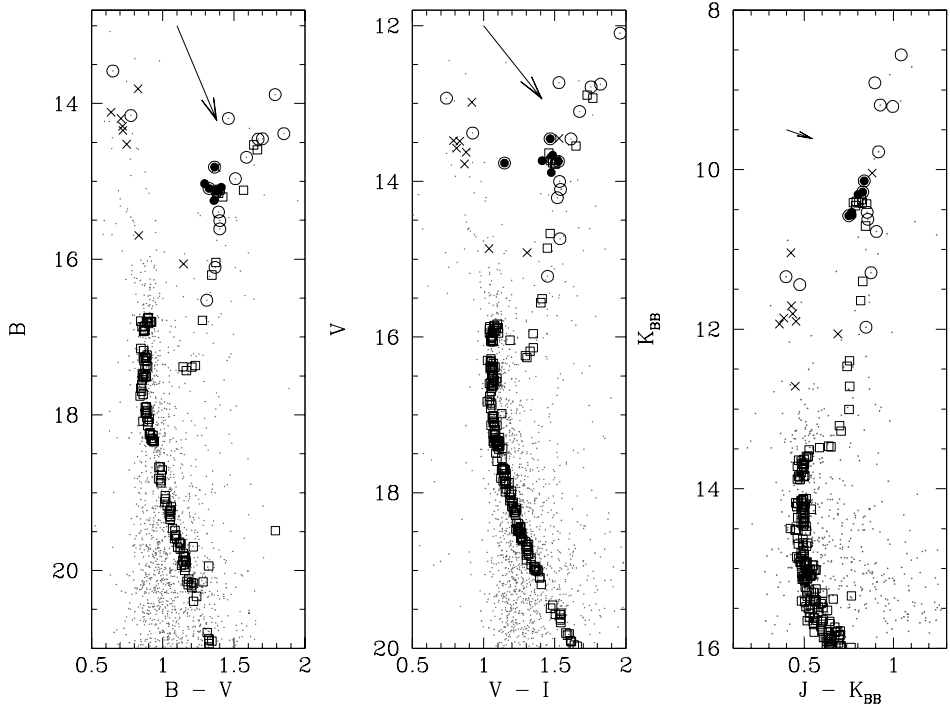}
\caption{Comparison of NGC 7142 photometry with likely single-star members of
  M67 ($\sq$; \citealt{sandm67}). Other symbols have the same meaning as in
  Fig. \ref{clumpcmd}. The photometry for M67 has been shifted in color to
  bring red clump stars into agreement in $V-I$, and the implied $E(V-I)$ was
  used to compute reddenings and extinctions in other filters according to the
  method of \citet{mccall}. Reddening vectors are shown in each diagram.
\label{m67comp}}
\end{figure}

Because photometry is available in $BVI_C$ \citep{sandm67} and $JHK_S$
\citep{2mass}, differences in median magnitudes of the clump of the clusters
in different filters allow a good estimate of the difference in distance
moduli. In the 2MASS filters where the effects of extinction are minimized,
these differences converge to about 2.4 mag. With a small correction for the
extinction, this implies $(m-M)_{0,N7142} - (m-M)_{0,M67} = 2.3$, and $(m-M)_0
= 11.9$. We can then look for consistency when the color-magnitude diagrams of
the clusters are overlaid in different filter combinations in order to check
the earlier estimate of the reddening. This requires that the extinctions and
reddenings in different filter combinations can be reliably related, which is
a complicated subject \citep{mccall}. For example, the measured reddening
should depend on the spectral type of the star being observed because the
flux-weighted wavelength of detected photons (the ``effective wavelength'')
changes.  The use of \citet{ccm} extinction relationships (including the use
of $R_V = A_V / E(B-V)$ values different from the canonical 3.1) did not
produce consistent CMDs in optical bands. In Fig. \ref{m67comp}, we show the
results of using the method of \citet{mccall} to compute reddenings and
extinctions from $E(V-I)$, which was itself estimated from the shifts between
the red clump stars in the two clusters and assuming the clump stars are
approximately K1-K2 III giants. Calculations were done using the York
Extinction Solver\footnote{\tt
  http://www2.cadc-ccda.hia-iha.nrc-cnrc.gc.ca/community/YorkExtinctionSolver/}. As
can be seen, the resulting CMDs show that the clumps, giant branch stars and
main sequence lines largely agree in position. 

The expected extinctions and
reddening for turnoff stars (near F5 V) only differ from the clump values by
0.02 (in the optical) at most. So we conclude that differential effects due to
the differing spectral energy distributions of giants and dwarfs are a minor
effect on the quality of the fit.

Based on the comparison with M67, NGC 7142 appears to be a slightly younger
cluster than M67. Even with some uncertainty in where to align the clumps of
the two clusters, there is a group of NGC 7142 stars that reach up to about
0.75 mag brighter than the turnoff of M67. Using measured values for M67 and
the results of the fit, our preferred values are $E(B-V) = 0.32 \pm 0.06$,
$(m-M)_0 = 11.9 \pm 0.15$, and $(m-M)_V = 12.96 \pm 0.24$.  The color
difference between the turnoff and the giant branch is similar for the two
clusters, although M67 has a larger difference in magnitude between the clump
and the turnoff, implying that NGC 7142 is younger.

\begin{figure}
% clusters/n7142/isos/cmdvsn6253.
\plotone{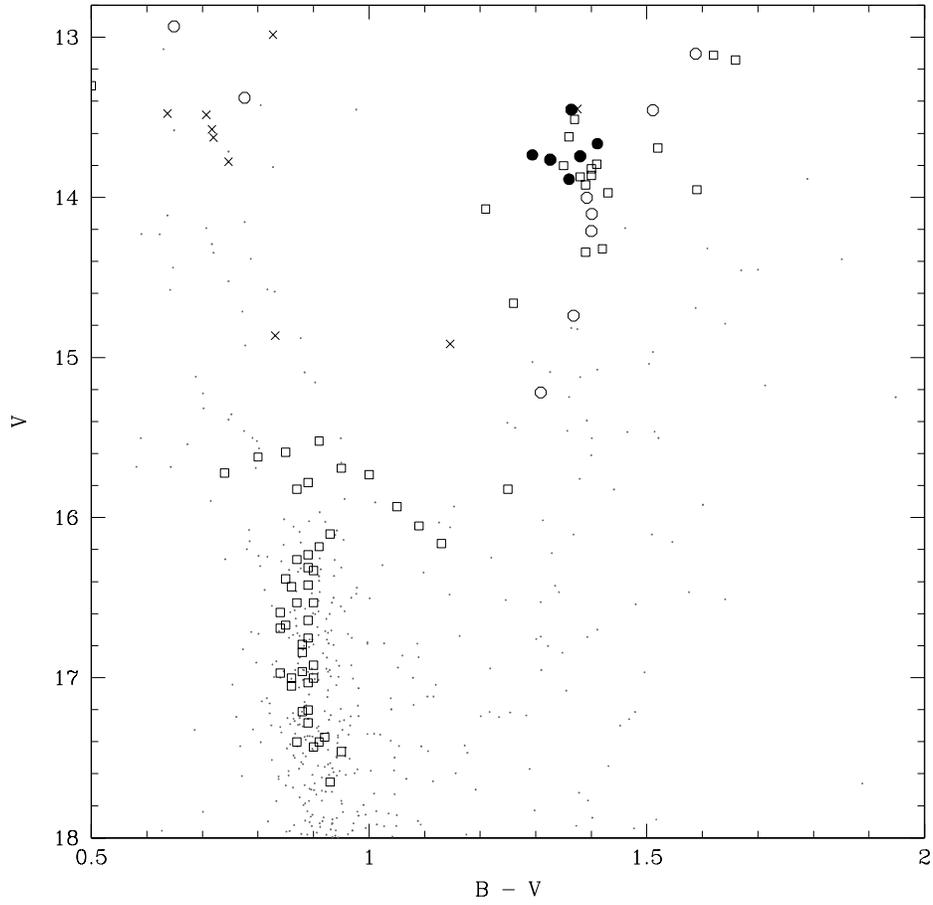}
\caption{Comparison of NGC 7142 photometry with likely single-star members of
  NGC 6253 ($\sq$; \citealt{twan6253}).  Other symbols have the same meaning
  as in Fig. \ref{clumpcmd}. The photometry for NGC 6253 has been shifted in
  color to bring red clump and giant stars into agreement.
\label{n6253comp}}
\end{figure}

At the suggestion of the referee, we also compared NGC 7142 to the more
metal-rich ([Fe/H] $=+0.43$; \citealt{atn6253}) open cluster NGC 6253. If the
NGC 6253 photometry is shifted [$\Delta V = 1.1$ mag, $\Delta (B-V) = 0.02$]
so that the red clumps are once again in approximate agreement, the main
sequences also agree as shown in Fig. \ref{n6253comp}. The brightness of
subgiant star candidates in NGC 6253 indicate that that cluster is similar in
age to or slightly younger than NGC 7142. Present indications are than NGC
6253 is around 3 Gyr old (see \citealt{mont} for a summary of determinations),
although the large metallicity of the cluster complicates its analysis.

\begin{figure}
% clusters/n7142/isos/cmdisos.bvi.
%\epsscale{1.0}
\plotone{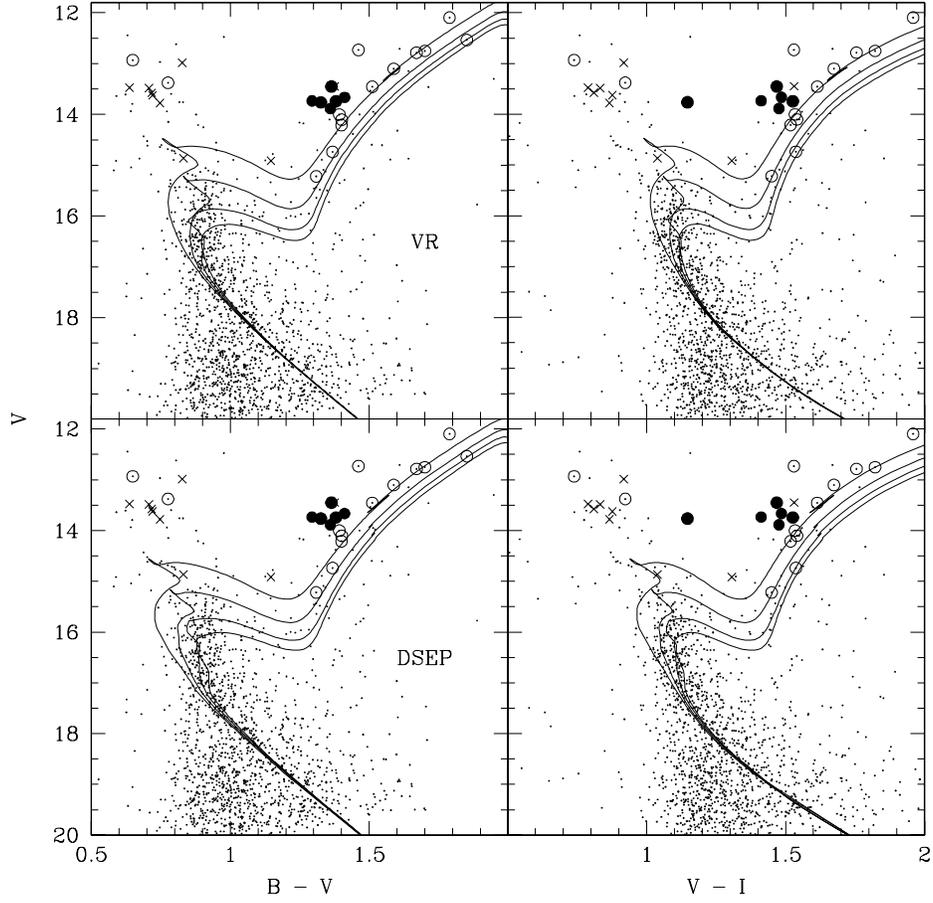}
\caption{A comparison of the $BVI$ CMDs with Victoria-Regina \citep{vr} and DSEP
  \citep{dsep} isochrones for 2, 3, 4, and 5 Gyr in age. In both cases, $(m-M)_0 =
  11.9$, $E(B-V) = 0.32$, $E(V-I) = 0.46$, and [Fe/H] $\approx +0.13$.
\label{isos1}}
\end{figure}

In Fig. \ref{isos1}, we present some illustrative comparisons with model
isochrones from two groups. Using distance and reddening derived from the
M67 comparison above and [Fe/H] equal to the most recent determination by
\citet{spec}, we find adequate matches to the main sequence, but imperfect
agreement with spectroscopically identified giants. Decreased [Fe/H] improves
the agreement with the giant branch, but does not eliminate it for reasonable
choices ([Fe/H] $> -0.1$). \citet{v2010}, for example, describe in detail
difficulties with the color-temperature transformations that must be used to
place theoretical models in the observational plane. For the Hyades, which
have [Fe/H] similar to NGC 7142, several transformation algorithms do a
reasonable job of reproducing the main sequence. However, when it comes to the
older and more metal-rich cluster NGC 6791, the authors found that the giant
branch was too red when \citet{vc03} transformations were used (as in both
sets of isochrones in Fig. \ref{isos1}). Given the current uncertainties in
the transformations and potential problems related to the composition of
super-metal rich stars (such as non-solar abundance ratios and helium
abundance), we will put off further discussion of the isochrones until full
analyses of the eclipsing binary stars is completed.

If the colors of main sequence stars are reliably modelled, the distance and
reddening derived from the red giant clump imply an age near 3 Gyr for the
cluster. Larger ages do not adequately model the bright end of the main
sequence or potential subgiant branch stars ($15.5 < V < 16$ and $1 < B-V <
1.3$). This is greatly different from the age of nearly 7 Gyr derived by
\cite{janes}, in spite of the fact that they derived a similar distance and
reddening. It may be that their technique (which involved fitting sub-solar
and solar-composition synthetic CMDs rather than super-solar ones) was
affected by some of the same mismatches on the giant branch we pointed out
above. If we tried to fit the color difference between the main sequence and
red giant branch at the level of the turnoff with current isochrones, a much
larger age would be obtained, but the red clump's magnitude would be poorly
matched.

% V-I issue is not photometry (Janes photometry also redder than isochrones)
% Vandenberg suggests V-I phot in his isochrones is too _red_ already by 0.02
% E(V-I)/E(B-V) has to be increased a lot to get consistency?

\section{Conclusions}

We have conducted an extensive variability search among stars in the field of
the cluster NGC 7142. Among the most important discoveries were several
detached eclipsing systems near the cluster turnoff (two of which have been
verified as cluster members based on radial velocities), a faint eclipsing
binary with shallow eclipses indicative of a small stellar companion, and a
long period variable found far to the red of the cluster giant branch.  The
detached eclipsing binaries allow us to put NGC 7142 on a growing list of
clusters with multiple double-lined binary stars that can provide masses and
radii for further tests of stellar models.

We have used the red giant clump in an attempt to derive a more reliable
distance and mean reddening for the cluster. The values derived here fall at
the low end of previous determinations for reddening, and at the high end for
distance. Our preferred values [$E(B-V) = 0.32 \pm 0.06$, $E(V-I)=0.46$,
  $(m-M)_0 = 11.9\pm0.15$] come from a fit of the 2MASS CMD of M67 to that of
NGC 7142. With this distance, models indicate that the cluster's age is near 3
Gyr. Future work is needed to check the agreement of theoretical models with
the CMD, however, as there are lingering difficulties with color-$T_{\rm eff}$
transformations and with our understanding of the compositions of
super-metal-rich stars.

NGC 7142 has characteristics putting it in an interesting regime. It appears
to be more metal rich than the Sun or the well-studied open cluster M67,
similar in metallicity to the Hyades, and more metal-poor than the very
metal-rich cluster NGC 6791. As such, it can help further test metal-rich
stellar models. NGC 7142's age also is interesting --- it is similar in age to
M67, making it a cluster with main sequence stars that are near the transition
between massive stars with convective cores and low-mass stars with radiative
cores. The main difficulties in deriving accurate cluster properties are the
fairly heavy reddening and field star contamination of the field. With future
membership surveys (radial velocity or proper motion) or with the use of
Stromgren photometry, these difficulties can be overcome.

\acknowledgments This work has been funded through grant AST 09-08536 from the
National Science Foundation to E.L.S. We would like to thank the Director of
Mount Laguna Observatory (P. Etzel) for generous allocations of observing
time. Infrastructure support for the observatory was generously provided by
the National Science Foundation through the Program for Research and Education
using Small Telescopes (PREST) under grant AST 05-19686. We would also like to
thank A. Bostroem, C. Gabler, and J. B. Leep for assisting with the
photometric calibration observations, and Mark Jeffries for taking some of the
later time-series photometry.

The Hobby-Eberly Telescope (HET) is a joint project of the University of Texas
at Austin, the Pennsylvania State University, Stanford University,
Ludwig-Maximilians-Universitat Munchen, and Georg-August-Universitat
Gottingen.  The HET is named in honor of its principal benefactors, William
P. Hobby and Robert E. Eberly.  This research made use of the SIMBAD database,
operated at CDS, Strasbourg, France; and the NASA/ IPAC Infrared Science
Archive, which is operated by the Jet Propulsion Laboratory, California
Institute of Technology, under contract with the National Aeronautics and
Space Administration.
%, and the WEBDA database, operated at the
%Institute for Astronomy of the University of Vienna.

%\begin{figure}
%% clusters/n7142/ew
%\plotone{lcv3.ps}
%\caption{$BVRI$ phased light curves for the near-contact binary V3.
%\label{lcv3}}
%\end{figure}

%\begin{figure}
%% clusters/n7142/ew
%\plotone{lcv9.ps}
%\caption{$BVRI$ phased light curves for the near-contact binary V9.
%\label{lcv9}}
%\end{figure}

%\begin{figure}
%% clusters/n7142/ew
%\plotone{lcv13.ps}
%\caption{$BVRI$ phased light curves for the near-contact binary V13.
%\label{lcv13}}
%\end{figure}

%\begin{figure}
%% clusters/n7142/2mass/cmd.
%\epsscale{0.8}
%\plotone{f16.ps}
%\caption{2MASS CMD for NGC 7142 with likely single-star members of M67 ($\sq$;
%  \citealt{sandm67}) overplotted. The photometry for M67 has been shifted
%  according to two possible differences in reddening and distance modulus
%  between the two clusters (see the text for further discussion of the choices).
%\label{m67comp}}
%\end{figure}

\clearpage
\begin{deluxetable}{clcl|clcl}
\tablewidth{0pt}
%\tabletypesize{\scriptsize}
\tablecaption{Photometry at Mount Laguna Observatory}
\tablehead{\colhead{Date} &
\colhead{Filters} & \colhead{mJD Start\tablenotemark{a}} & \colhead{$N$}
& \colhead{Date} &
\colhead{Filters} & \colhead{mJD Start\tablenotemark{a}} & \colhead{$N$}}
\startdata
Aug. 10, 2005 & $R$ & 3593.755 & 44 & Jul. 7, 2008 & $I_c$ & 4655.698 & 158 \\
Aug. 11, 2005 & $R$ & 3594.687 & 39 & Jul. 8, 2008 & $I_c$ & 4656.686 & 204 \\ 
Aug. 13, 2005 & $R$ & 3596.670 & 46 & Jul. 9, 2008 & $I_c$ & 4657.684 & 96 \\
Aug. 14, 2005 & $R$ & 3597.659 & 19 & Jul. 27, 2008 & $V$ & 4675.660 & 97 \\
Aug. 15, 2005 & $R$ & 3598.642 & 50 & Jul. 28, 2008 & $V$ & 4676.655 & 103 \\ 
Aug. 16, 2005 & $R$ & 3599.666 & 42 & Jul. 29, 2008 & $V$ & 4677.658 & 105 \\ 
Aug. 17, 2005 & $R$ & 3600.659 & 48 & Jul. 31, 2008 & $V$ & 4678.653 & 107 \\ 
Sep. 21, 2005 & $R$ & 3635.749 & 40 & Aug. 18, 2008 & $V$ & 4697.952 & 34 \\   
Sep. 22, 2005 & $R$ & 3636.607 & 32 & Oct. 6, 2008 & $VR_K$ & 4746.674 & 11,10 \\
Sep. 23, 2005 & $R$ & 3637.613 & 68 & Oct. 25, 2008 & $BVI_C$ & 4765.686 & 9,8,11 \\
Sep. 24, 2005 & $R$ & 3638.612 & 30 & Sep. 6, 2009 & $V$ & 5081.895 & 31 \\  
Sep. 25, 2005 & $R$ & 3639.605 & 68 & Oct. 9, 2009 & $V$ & 5114.592 & 41 \\  
Sep. 26, 2005 & $R$ & 3640.704 & 28 & Oct. 16, 2009 & $V$ & 5121.578 & 66 \\ 
Sep. 27, 2005 & $R$ & 3641.608 & 48 & Nov. 10, 2009 & $V$ & 5146.575 & 52 \\
Oct. 6, 2005 & $R$ & 3650.607 & 38 & Nov. 29, 2009 & $V$ & 5165.590 & 13 \\ 
Oct. 7, 2005 & $R$ & 3651.608 & 46 & May 19, 2010 & $I_C$ & 5336.815 & 37 \\
Oct. 8, 2005 & $R$ & 3652.597 & 33 & Jun. 7, 2010 & $B$ & 5355.742 & 55\\   
Oct. 9, 2005 & $BVR$ & 3653.600 & 11,11,11 &  Jun. 24, 2010 & $V$ & 5372.709 & 68 \\
Nov. 12, 2007 & $BVR_K$ & 4417.703 & 11,16,10 & Jul. 5, 2010 & $R$ & 5383.674 & 76 \\   
Nov. 14, 2007 & $BVR_K$ & 4419.612 & 13,11,11 & Jul. 12, 2010 & $I_C$ & 5390.669 & 75 \\
Jun. 8, 2008 & $B$ & 4626.780 & 55 & Aug. 2, 2010 & $VI_C$ & 5411.665 & 12,61 \\
Jun. 9, 2008 & $B$ & 4627.796 & 50 & Aug. 3, 2010 & $I_C$ & 5412.660 & 11 \\
Jun. 11, 2008 & $B$ & 4629.764 & 57 & Aug. 14, 2010 & $V$ & 5423.686 & 18 \\
Jun. 12, 2008 & $B$ & 4630.753 & 60 & Nov. 16, 2010 & $VI_C$ & 5517.570 & 70,5 \\
Jun. 15, 2008 & $B$ & 4633.746 & 60 & Jun. 23, 2011 & $I_C$ & 5736.721 & 48 \\ 
Jun. 16, 2008 & $B$ & 4634.747 & 57 & Jul. 12, 2011 & $RI_C$ & 5755.670 & 16,65 \\
Jun. 17, 2008 & $I_c$ & 4635.740 & 159 \\
\enddata
\label{phottab}
\tablenotetext{a}{mJD = HJD - 2450000.}  
\end{deluxetable}

\begin{deluxetable}{lccr}
\tablewidth{0pt}
%\tabletypesize{\scriptsize}
\tablecaption{Systematic Radial Velocities}
\tablehead{\colhead{ID\tablenotemark{a}} & \colhead{$v_{CoM}$ (\kms)} & \colhead{$\sigma$ (\kms)} & \colhead{$N_{obs}$}}
\startdata
\multicolumn{4}{l}{Variable Stars:}\\
\hline
V1 & $-17.0$ & 1.0 & 3\\
V2 & $-42.1$ & 0.6 & 13 \\%-42.1 \pm 0.1 from chi^2
V375 Cep & $-49.1$ & 1.7 & 21 \\%-48.8 \pm 0.5 from chi^2
\multicolumn{4}{l}{Red Clump Stars:}\\
\hline
JH2222& $-49.6$ & 0.1 & 1 \\
JH2288& $-44.0$ & 0.1 & 1 \\
JH1003& $-43.9$ & 0.1 & 1 \\
\enddata
\tablenotetext{a}{V: Variable star identifier, JH: identification number from \citet{janes}.}
\label{spectab}
\end{deluxetable}

% my IDs from n7142cal.dat
% photometry has been corrected to times of maximum light where possible
\begin{deluxetable}{rrcllllllll}
\tablewidth{0pt}
\tabletypesize{\scriptsize}
\tablecaption{Variable Stars Detected in the NGC 7142 Field}
\tablehead{& \colhead{CT\tablenotemark{a}}& \colhead{Type\tablenotemark{b}} & \colhead{RA} & \colhead{DEC} &
\colhead{$V$} &
\colhead{$B-V$} & \colhead{$V-I$} & \colhead{$P$ (d)} & \colhead{$E_0$} &
\colhead{Notes\tablenotemark{c}}} 
\startdata
% 490; lc530 - reexamine later - 1 possible eclipse in R
V1 & 395 & EA & 21:44:32.86 & +65:45:26.3 & 14.864 & 0.831 & 1.039 & 4.6691 & 2453639.93 & rv prob. nm\\ %lc102, originally ID83
V2 & 18 & EA & 21:44:29.62 & +65:48:43.9 & 15.310 & 0.797 & 1.010 & 15.6505 & 2453639.684 & deep eclipses; eccentric; rv mem\\ %lc138 new in I,V data; original ID 116; J=13.661+\-0.032 Ks=13.227+\-0.037
V3 & 155 & EW & 21:45:15.16 & +65:49:24.3 & 15.38 & 0.72 & 0.93 & 0.580793 & 2453594.80 & straggler? \\ % lc163; Serio ID 2146; original ID133
V4 & 430 & LPV & 21:44:55.98 & +65:45:50.0 & 15.43 & 1.89 & 2.06 & & & VH Y\\ % lc108; new in I,V data; long period pulsator?; or something weirder?; original ID123
V5 & 199? & EW & 21:45:30.03 & +65:46:42.3 & 15.80 & 1.18 & 1.38 & 0.3368235 & 2453596.782 & nearby faint star\\ %lc555; Serio ID 2703; not marked in CT; phot consistent; original ID 191
V6 & & EB & 21:44:13.21 & +65:45:01.3 & 15.71 & 0.91 & 1.15 & 0.441163 & 2453599.757 & \\ % lc214; Serio ID 109; outside CT field; original ID189; B = 16.62 V = 15.71 I = 14.56
V7 & 288 & EW & 21:45:10.77 & +65:44:41.3 & 15.95 & 0.89 & 1.12 & 0.69531? & & VH f\\ % lc236; Serio ID 1978; original ID209
   & & EA & 21:44:54.74 & +65:43:55.3 & 16.115 & 0.877 & 1.080 & 1.909682 & 2454676.8071 & V375 Cep; rv mem\\ %Serio ID 1418, lc272; original ID249
%   & & EA & 21:44:54.7 & +65:43:55.3 & 16.115 & 0.877 & 1.080 & 1.909685 & 2453599.748 & V375 Cep; rv mem\\ %Serio ID 1418, lc272; original ID249
V8 & & EA & 21:44:35.73 & +65:45:04.7 & 17.288 & 1.106 & 1.373 & 0.52730 & 2453596.985 & triple? \\ % lc549; Serio ID 799; not IDd in CT; original ID 504
V9 & & EW & 21:44:28.44 & +65:46:36.6 & 17.63 & 1.14 & 1.34 & 0.330233 & 2453593.918 & \\ % lc682; Serio ID 589; not IDd in CT, original ID628
V10 & 356 & Irr & 21:45:40.73 & +65:43:40.0 & 17.63 & 1.25 & 1.53 &  & & \\ % real? lc638; Serio ID 3028; original ID 601
V11 & & EA & 21:44:17.92 & +65:50:04.9 & 17.906 & 1.055 & 1.304 & 1.28629 & 2453598.825 & \\ % lc761; Serio ID 254; outside CT field; original ID689
V12 & & EW? & 21:44:55.46 & +65:52:59.0 & 18.330 & 1.157 & 1.340 & 0.2938 & 2453636.79 & \\ % lc898; Serio ID 1438; outside CT field; original ID852
V13 & & EW & 21:44:20.66 & +65:44:59.1 & 18.35 & 1.08 & 1.20 & 0.34207 & 2453598.89 & total eclipses, prob. nm\\ % lc977; Serio ID 345; original ID916
V14 & & EW & 21:46:06.8 & +65:52:53.6 & & & & 0.29005 & 2453600.75 & not detected in calibrating obs.\\ % Serio ID 3796; faint; outside CT field
%148 & LPV & 21:45:12.8 & +65:47:02.9 & 15.577 & 1.163 & 1.106 & & & CT159\\ % real? lc152; long period? not real convincing in BVI
% 82 &  & 14.870 & 1.641 & 1.854 & & CT303\\ % lc73; Serio ID 2242; not real?
% 70 & $\delta$ Scu? &  &  &  & & \\ % needs confirmation
% 167; lc94 is near V4 in CMD, but is NOT varying...
\enddata \tablenotetext{a}{CT: Identification number from \citet{ct}}
\tablenotetext{b}{EW: eclipsing contact or near-contact binary; EB: detached
  eclipsing binary with stars having ellipsoidal distortions; EA:
  well-detached eclipsing binary; LPV: long period variable star; Irr:
  irregular brightness variations} \tablenotetext{c}{nm: nonmember; rv mem:
  membership inferred from radial velocities}
\label{vartab}
\end{deluxetable}

\begin{deluxetable}{rrcccccccl}
\tablewidth{0pt} \tabletypesize{\scriptsize} \tablecaption{Red Clump Star
  Candidates in the NGC 7142 Field} \tablehead{\colhead{CT
    ID\tablenotemark{a}}&\colhead{JH ID\tablenotemark{a}}& \colhead{RA} &
  \colhead{DEC} & \colhead{$V$} & \colhead{$B-V$} & \colhead{$V-I$} &
  \colhead{$K_s$} & \colhead{$(J-K_S)$} & \colhead{Notes\tablenotemark{b}}}
\startdata
% Janes center: 21:45:10  65:46:18
\multicolumn{10}{l}{Probable Clump Stars:}\\
421 & 1065 & 21:45:00.73 & +65:45:56.5 & 13.452 & 1.364 & 1.468 & 10.097 & 0.802 & I; rv, abund mem\\ % ID 26
170 & 1519 & 21:45:20.43 & +65:48:31.0 & 13.743 & 1.380 & 1.526 & 10.237 & 0.794 & O,rv mem?\\ %ID 37
203 & 1767 & 21:45:32.93 & +65:47:31.1 & 13.765 & 1.326 & 1.147 & 10.533 & 0.718 & rv, abund mem;nearby star affecting phot? \\ % ID 39
& 2222 & 21:45:57.03 & +65:44:10.5 & 13.665 & 1.411 & 1.485 & 10.268 & 0.768 & rv mem,off CT field\\ %ID 35
93 & 1003 & 21:44:57.60 & +65:48:36.8 & 13.735 & 1.294 & 1.412 & 10.490 & 0.734 & rv mem?\\ % ID 38
 & 2288 & 21:46:00.93 & +65:49:06.8 & 13.887 & 1.360 & 1.476 & 10.526 & 0.736 & rv mem?,off CT field\\ %ID 43
%\multicolumn{10}{l}{Possible Giant Stars:}\\
% 41 & & 21:44:38.82 & +65:46:38.2 & 14.002 & 1.392 & 1.534 & 10.489 & 0.818 & 120,rv mem?\\ % ID 48
%    & & 21:46:08.58 & +65:54:58.9 & 14.101 & 1.357 & 1.549 & 10.632 & 0.814 & off CT field\\ %ID 52
% 51 & & 21:44:43.82 & +65:46:42.5 & 14.103 & 1.401 & 1.541 & 10.579 & 0.822 & \\ %ID 53
\multicolumn{10}{l}{Probable nonmember:}\\
399 & 591 & 21:44:36.70 & +65:43:19.1 & 13.448 & 1.375 & 1.531 &  9.998 & 0.845 & rv nm\\ %ID 24
\enddata \tablenotetext{a}{CT: Identification number from \citet{ct}, JH: identification number from \citet{janes}.}
\tablenotetext{b}{nm: nonmember; mem: member; rv: inference from radial
  velocities; abund: inference from abundances}
\label{rctab}
\end{deluxetable}

\begin{deluxetable}{lr|lr|lr}
\tablewidth{0pt} 
\tablecaption{Estimated Extinction and Reddening Values relative to M67}
\tablehead{\colhead{Quantity}&\colhead{Value}& \colhead{Quantity} &
  \colhead{Value}&
 \colhead{Quantity} &  \colhead{Value}}
\startdata
$\Delta E(V-I)$ & 0.41 & $\Delta \tau_1$\tablenotemark{a} & 0.335 & & \\
$\Delta A_B$ & 1.22 & $\Delta (B-V)$ & 0.28 & $A_V / E(B-V)$ & 3.33 \\
$\Delta A_V$ & 0.94 & & & $A_V / E(V-I)$ & 2.30\\
$\Delta A_{K_s}$ & 0.11 & $\Delta E(J-K_s)$ & 0.15 & $A_{K_s} / E(J-K_s)$ & 0.76\\
\enddata
\tablenotetext{a}{$\tau_1$ is the optical depth at 1 $\mu$m.}
\label{exttab}
\end{deluxetable}

\end{document}